\documentclass[sn-nature]{sn-jnl}

\usepackage{graphicx}%
\usepackage{multirow}%
\usepackage{amsmath,amssymb,amsfonts}%
\usepackage{amsthm}%
\usepackage{mathrsfs}%
\usepackage[title]{appendix}%
\usepackage{xcolor}%
\usepackage{textcomp}%
\usepackage{manyfoot}%
\usepackage{booktabs}%
\usepackage{algorithm}%
\usepackage{algorithmicx}%
\usepackage{algpseudocode}%
\usepackage{listings}%
\usepackage{siunitx}%

\DeclareUnicodeCharacter{2212}{-}
\DeclareSIUnit\angstrom{\text {Å}}

\begin{document}

\title{Advancing Nonadiabatic Molecular Dynamics Simulations for Solids: Achieving Supreme Accuracy and Efficiency with Machine Learning}

\author[1]{\fnm{Changwei} \sur{Zhang}}
\equalcont{These authors contributed equally to this work.}
\author[1]{\fnm{Yang} \sur{Zhong}}
\equalcont{These authors contributed equally to this work.}

\author[1]{\fnm{Zhi-Guo} \sur{Tao}}

\author[2]{\fnm{Xinming} \sur{Qing}}

\author[2]{\fnm{Honghui} \sur{Shang}}
\author[3]{\fnm{Zhenggang} \sur{Lan}}
\author[4]{\fnm{Oleg V.}\sur{Prezhdo}}

\author[1]{\fnm{Xin-Gao} \sur{Gong}}

\author*[1]{\fnm{Weibin} \sur{Chu}}\email{wbchu@fudan.edu.cn}

\author*[1]{\fnm{Hongjun} \sur{Xiang}}\email{hxiang@fudan.edu.cn}

\affil[1]{\orgdiv{Key Laboratory of Computational Physical Sciences (Ministry of Education), Institute of Computational Physical Sciences, State Key Laboratory of Surface Physics, and Department of Physics}, \orgname{Fudan University},\orgaddress{\city{ Shanghai}, \postcode{200433}, \country{China}}}

\affil[2]{\orgdiv{Key Laboratory of Precision and Intelligent Chemistry, Hefei National Research Center for Physical Sciences at the Microscale}, \orgname{University of Science and Technology of China}, \orgaddress{\city{Hefei, Anhui}, \postcode{230026}, \country{China}}}

\affil[3]{\orgdiv{Environmental Research Institute, Guangdong Provincial Key Laboratory of Chemical 
Pollution and Environmental Safety \& MOE Key Laboratory of Environmental Theoretical 
Chemistry}, \orgname{South China Normal University}, \orgaddress{\city{Guangzhou, Guangdong}, \postcode{510006}, \country{China}}}

\affil[4]{\orgdiv{Department of Chemistry and Department of Physics \& Astronomy}, \orgname{University of Southern California}, \orgaddress{\city{Los Angeles California}, \postcode{90089}, \country{United States}}}

\abstract{
Non-adiabatic molecular dynamics (NAMD) simulations have become an indispensable tool for investigating excited-state dynamics in solids. In this work, we propose a general framework, N\textsuperscript{2}AMD, which employs an E(3)-equivariant deep neural Hamiltonian to boost the accuracy and efficiency of NAMD simulations. The preservation of Euclidean symmetry of Hamiltonian enables N\textsuperscript{2}AMD to achieve state-of-the-art performance. Distinct from conventional machine learning methods that predict key quantities in NAMD, N\textsuperscript{2}AMD computes these quantities directly with a deep neural Hamiltonian, ensuring supreme accuracy, efficiency, and consistency. Furthermore, N\textsuperscript{2}AMD demonstrates excellent generalizability and enables seamless integration with advanced NAMD techniques and infrastructures. Taking several extensively investigated semiconductors as the prototypical system, we successfully simulate carrier recombination in both pristine and defective systems at large scales where conventional NAMD often significantly underestimates or even qualitatively incorrectly predicts lifetimes. This framework not only boosts the efficiency and precision of NAMD simulations but also opens new avenues to advance materials research.
}

\maketitle

\section{\label{introduction}Introduction}


In recent years, Non-adiabatic molecular dynamics (NAMD) has achieved remarkable success in revealing the ultrafast microscopic mechanism of excited state dynamics in complex systems where electronic and nuclear dynamics are strongly coupled\cite{zheng_ab_2023, zheng_Multiple_2023, nelson_non-adiabatic_2020, zheng_ab_2019}. This methodology proves indispensable across various fields, including photovoltaics photocatalysis, and optoelectronics, where it plays a crucial role in elucidating energy conversion processes\cite{long_Nonadiabatic_2017, zheng_phonon-assisted_2017, lian_ultrafast_2020, chu_ultrafast_2022, prezhdo_modeling_2021}. Particularly in energy conversion devices such as solar cells and light-emitting diodes, understanding and managing nonradiative electron-hole recombination is vital for enhancing device efficiency and performance.

However, the efficiency and accuracy of NAMD simulations are significantly inferior compared to ground-state calculations. The computational demands of NAMD are several orders of magnitude higher than those for static calculations. Furthermore, the accuracy of NAMD simulations is strongly dependent on the choice of exchange-correlation functional used in electronic structure calculations, owing to the crucial role of the energy differences between molecular orbitals and non-adiabatic couplings (NAC) \cite{lin_dependence_2016, zhu_density_2021}. Particularly in predicting nonradiative electron-hole recombination, the commonly used Local Density Approximation (LDA) or Generalized Gradient Approximation (GGA) for exchange-correlation often suffers from the notorious self-interaction error, which can even lead to qualitatively incorrect results.

Efforts to enhance the reliability of NAMD simulations are ongoing. One notable strategy involves the DFT+U method\cite{Anisimov_band_1991}, which introduces a Hubbard U parameter to account for the Coulombic repulsion among multiple electrons occupying the same site, particularly improving band gap estimations. However, selecting an optimal U parameter remains challenging\cite{hu_choice_2011, loschen_First-principles_2007}, and its use is limited to systems with localized electrons, even though the underestimation of band gaps is a widespread issue for all systems. Another approach in NAMD is employing the scissor operation\cite{wang_real_2019}, which effectively adjusts energy levels but does not modify band dispersion, the time derivative of band energy, or wavefunctions, leaving the correction of NAC unresolved.

Thus, NAMD simulations employing hybrid functionals offer a more robust solution compared to conventional functionals and correction methods. However, the computational costs associated with hybrid functional calculations are substantially higher than those for local and semi-local functionals. NAMD involves the real-time evolution of the time-dependent Schrödinger equation, primarily requiring repeated solutions of the electronic structure—often tens of thousands of times—to capture NAC at each time step. Consequently, the application of hybrid functionals in NAMD generally becomes infeasible for solid-state materials due to the high computational demand.

The rise of machine learning (ML) offers a promising avenue for accelerating NAMD simulations, with significant efforts dedicated to using ML to accelerate the calculation of NAC\cite{Cignoni_machine_2023, wang_all-atom_2022, dral_Molecular_2021, zhang_mlatom_2024}. Tretiak and colleagues developed a hierarchically interacting particle neural network to predict non-adiabatic coupling vectors (NACVs), which was subsequently applied to compute exciton polaron properties in azomethanes\cite{li_machine_2024} and plasmon dynamics\cite{habib_machine_2023}. Given the notable challenge of directly predicting NAC, alternative frameworks have been proposed. One approach integrates ML with generalization of the Landau–Zener algorithm\cite{hu_inclusion_2018, zhang_mlatom_2024}, where NAC is not present in real-time propagation. Another method approximates NAC using the Baeck-An scheme\cite{shu_Nonadiabatic_2022, T_do_Casal_fewest_2021}. For instance, Marquetand and co-workers\cite{Westermayr_combining_2020} combined SchNet and SHARC to perform NAMD with ML potentials, their gradients, and Hessian. Lopez and co-workers\cite{li_Automatic_2021} developed $\mathrm{PyRAI^2MD}$ and used it to investigate the photoisomerization mechanism. Recently, inspired by successful predictions of molecular Hamiltonian matrices, evaluating NAC with ML Hamiltonians has shown great potential in ML-NAMD. Akimov and colleagues\cite{Shakiba_Machine-Learned_2024} employed KRR to map between non-self-consistent and self-consistent Hamiltonians calculated via different functionals, providing deeper insights into excitation energy relaxation in C$_{60}$ fullerene and Si$_{75}$H$_{64}$ at reduced computational costs.

Despite all these achievements, the accuracy and transferability of ML methods in excited state dynamics remain significantly lower compared to their performance in ground state dynamics, restricting their widespread application in NAMD simulations. This issue is particularly pronounced in solid-state systems, where the chemical environments of atoms are considerably more complex due to periodic boundary conditions.  In many cases, earlier work could only predict qualitative results\cite{zhang_Doping-Induced_2021}. To address these limitations, Prezhdo and co-workers proposed using ML to interpolate the NAC along an MD trajectory for solids\cite{wang_interpolating_2023}, which can significantly reduce computational costs. However, this framework lacks the ability to extrapolate or predict NAC for novel configurations outside the training set. This highlights the ongoing challenge of developing ML models that can accurately generalize to diverse excited state energy landscapes in solids. Recently, E(3) equivariant graph neural network (GNN) has been proven to be the state-of-the-art architecture for representing the mapping from structure to interatomic force field (FF) and DFT Hamiltonian\cite{gong_general_2023, Musaelian_learning_2023, zhong_transferable_2023, zhong_Universal_2024, tang_efficient_2023}. Utilizing these advanced GNN models in NAMD could substantially enhance both simulation accuracy and generalization capabilities while maintaining competitive computational costs.

In this work, we propose a general NAMD workflow augmented by E(3)-equivariant ML models, $\mathrm{N^2AMD}$ (Neural-Network Non-Adiabatic Molecular Dynamics), which enables efficient NAMD simulation of large-scale materials at the hybrid-functional level accuracy. We demonstrate the effectiveness of $\mathrm{N^2AMD}$ using several extensively studied semiconductors: Rutile Titanium Dioxide ($\mathrm{TiO_2}$), Gallium Arsenide ($\mathrm{GaAs}$), Molybdenum Disulfide ($\mathrm{MoS_2}$) and Silicon. Our systematic investigation of nonradiative recombination illustrates the state-of-the-art performance of $\mathrm{N^2AMD}$ in both pristine and defective systems, where conventional NAMD simulations typically fail. Conventional NAMD simulations using the Perdew-Burke-Ernzerhof (PBE) functional\cite{perdew_generalized_1996} severely underestimate the timescale by a factor of 10. This underestimation persists even when employing the widely used scissors correction. Furthermore, $\mathrm{N^2AMD}$ shows an extended capability to predict NAC vectors which are crucial for advancing beyond the current implementations of NAMD in solids. We anticipate that the proposed framework can be generally integrated with recently developed advanced NAMD methodologies, potentially revolutionizing developments in NAMD and fertilizing future materials research in nanoscale and condensed matter systems.

\section{\label{result}Results}
\subsection{\label{results21} Theoretical framework of $\mathrm{N^2AMD}$}

In the NAMD approach, the evolution of charge carriers in coupled electronic and nuclear dynamics is described by the time-dependent Schrödinger equation (TDSE):
\begin{equation}
    i\hbar \frac{\partial}{\partial t} \Psi(\mathbf{r}, \mathbf{R}, t) = \hat{H}(\mathbf{r}, \mathbf{R}) \Psi(\mathbf{r}, \mathbf{R}, t)
\label{eq1a}
\end{equation}
where $\mathbf{r}$ and $\mathbf{R}$ are the collective coordinates of the electrons and nuclei, respectively. And $\hat{H}(\mathbf{r}, \mathbf{R}) = \hat{T}(\mathbf{R}) + \hat{H}_{el}(\mathbf{r}, \mathbf{R})$, is the total Hamiltonian including the kinetic energy operators $\hat{T}(\mathbf{R})$, and electronic hamiltonian operator $\hat{H}_{el}(\mathbf{r}, \mathbf{R})$.

The most computationally efficient method for incorporating nonadiabatic effects is through mixed quantum-classical approaches, where nuclei are treated as classical particles and electrons are described quantum mechanically. Therefore, the TDSE is reduced to describe the electronic subsystem.

By representing the electronic wavefunction as a linear combination of instantaneous adiabatic Kohn-Sham orbitals functions $\{\psi_i\}$:
\begin{equation}
    \Psi(\mathbf{r}, \mathbf{R}, t) = \sum_i c_i(t) \psi_i(\mathbf{r}, \mathbf{R}(t))
\label{eq1b}
\end{equation}
The TDSE, eq~\ref{eq1a} can be reduced to a set of coupled differential equations for $c_i(t)$ coefficients:
\begin{equation}
    i\hbar \dot{c}_i(t) = \sum_j c_j(t) \left( E_j \delta_{ij} -i\hbar \mathbf{d}_{ij} \cdot \dot{\mathbf{R}}
    \right)
\label{eq1c}
\end{equation}
where $E_j$ is the energy of the $j$th adiabatic Kohn-Sham state, $\mathbf{d}_{ij} = \left< \psi_i | \nabla_R | \psi_j \right>$ is the NACV between state $i$ and state $j$, and $\dot{\mathbf{R}}$ is the velocity of nuclei. Typically, computing NACVs in complex systems such as solids is not feasible due to the computational cost. However, by employing Leibniz's notation of chain rule, the product of nonadiabatic coupling and nuclear velocity can be transformed into a time derivative:
\begin{equation}
    d_{ij} = \mathbf{d}_{ij} \cdot \dot{\mathbf{R}} = \left< \psi_i \left| \frac{\partial}{\partial t} \right| \psi_j\right>
\label{eq1d}
\end{equation}
where $d_{ij}$ is the so-called nonadiabatic coupling scalar
and it can be further numerically calculated using the Hammes-Schiffer-Tully formula\cite{hammes-schiffer_proton_1994}:
\begin{equation}
    d_{ij}\left(t+\frac{1}{2}dt\right) = \frac{\left< \psi_i(t)|\psi_j(t+dt) \right> - \left< \psi_i(t+dt)|\psi_j(t) \right>}{2dt}
\label{eq1e}
\end{equation}

Following eq~\ref{eq1c}, there are two prominent approaches for propagating coupled nuclear dynamics: Ehrenfest Dynamics and Trajectory Surface Hopping (TSH). In the Ehrenfest approach, nuclei move classically on an average potential energy surface. Conversely, in the TSH approach, nuclei move on a single adiabatic potential energy surface at a time, with stochastic "hops" between surfaces permitted. Considering our focus on the dynamics of non-equilibrium charge carriers in solids, especially non-radiative recombination processes that can last up to nanoseconds, our analysis will primarily focus on the TSH approach. This method facilitates a more straightforward accounting of decoherence and detailed balance. It should be noted that the proposed $\mathrm{N^2AMD}$ model is not limited to the TSH approach but can be generally applied to Ehrenfest Dynamics as well.

The Fewest Switches Surface Hopping (FSSH) method\cite{tully_molecular_1990}, proposed by Tully, is the most widespread TSH approach for simulations in both molecules and solids. To accurately simulate non-radiative recombination processes, it is essential to incorporate decoherence corrections into surface-hopping algorithms. For this purpose, we have utilized the Decoherence Induced Surface Hopping (DISH) method here\cite{jaeger_Decoherence-induced_2012}.

In the NAMD simulation of periodic solid-state materials, the TSH method is further simplified by utilizing the classical path approximation (CPA)\cite{akimov_pyxaid_2013}, where the trajectory $\mathbf{R}(t) $ is obtained by Bohn-Oppenheimer MD. CPA is a powerful approximation that significantly reduces computational complexity. Its effectiveness has been validated for solids under conditions where the presence of excited carriers does not induce significant lattice permutations or reforming during the dynamics\cite{prezhdo_modeling_2021, wang_mixed_2011}.

\subsection{\label{results22} Neural network architecture of $\mathrm{N^2AMD}$}
\begin{figure}[t]
\centering
\includegraphics[width=0.9\linewidth]{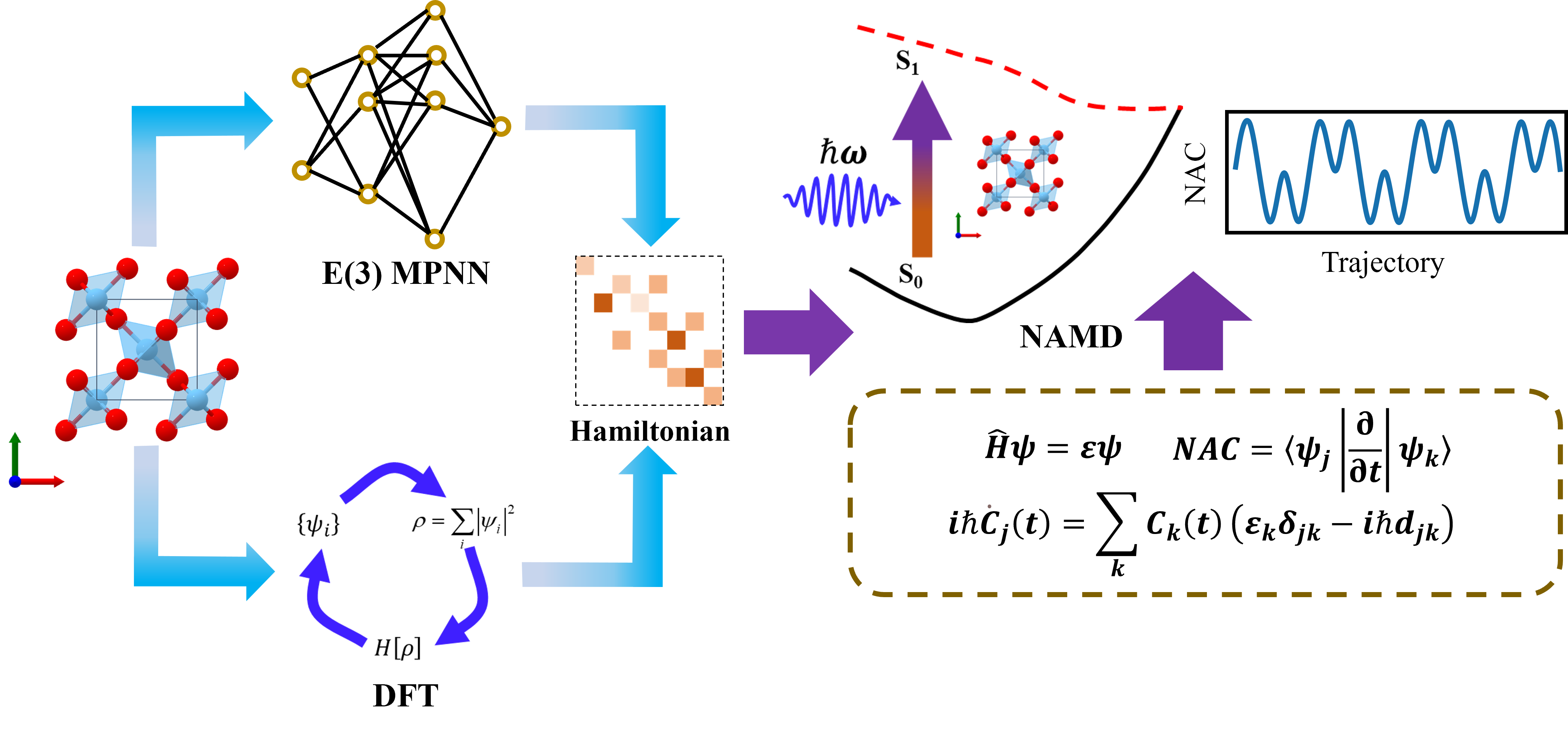}
\caption{\label{fig1} Schematic workflow of $\mathrm{N^2AMD}$. 
The graph neural network directly maps from crystal structures to instantaneous Hamiltonian matrices, bypassing the self-consistent iterative procedure in DFT. From the predicted Hamiltonian, $\mathrm{N^2AMD}$ evaluates critical quantities necessary for NAMD, such as KS orbital energies and NACs. These quantities are then utilized to perform excited dynamics simulations.
}
\end{figure}

The primary challenge in implementing NAMD with DFT lies in the computation of several key quantities in eq~\ref{eq1c}, which is hindered by high computational costs and the need for advanced functionals to ensure accuracy in electronic calculation. Therefore, we propose a general framework that utilizes the recently developed E(3) equivariant graph neural network to efficiently and accurately compute these quantities. As depicted in Figure~\ref{fig1}, this framework constructs the instantaneous Hamiltonian matrix in real space by mapping the on-site Hamiltonian and the off-site Hamiltonian matrix from the node features and edge features in the crystal structure. The detailed description of the neural network architecture is discussed in Ref.\cite{zhong_transferable_2023}. After transforming the real-space Hamiltonian to k-space, the instantaneous adiabatic basis functions in eq~\ref{eq1b} can be obtained by diagonalizing the Hamiltonian matrix. The corresponding eigenvalues in eq~\ref{eq1c} can also be determined through diagonalization. 
Given that $\mathrm{N^2AMD}$ utilizes a numerical atomic orbital (NAO) basis where the basis functions are not orthogonal, the NAC used in eq~\ref{eq1e} should be computed as:
\begin{equation}
    \left< \psi_i(t) | \psi_j(t+dt) \right> = \sum_{ab} \phi_{ia}^*(t) S_{ab}^{(k=0)}(t; t+dt) \phi_{jb}(t+dt)
\label{eq1h}
\end{equation}
where $\phi_{ia}$ is the $a$th vector component of the $i$th NAO basis function, and $S_{ab}^{(k)}$ is the overlap matrices, which can be obtained by Fourier transforming from real-space tight binding overlap matrices:
\begin{equation}
    S_{ab}^{(k)} = \sum_n e^{i\mathbf{k}\cdot \mathbf{R}_n} S_{ab}^{(R)}
\end{equation}

The accuracy of these critical quantities in NAMD is guaranteed through the use of E(3) equivariant mapping. The electronic Hamiltonian matrix represented in the atomic orbital basis must satisfy two key symmetry constraints: rotational equivariance and parity symmetry. For crystalline solids, the Hamiltonian also exhibits translational invariance. These fundamental symmetries belong to the E(3) group, which includes rotations, translations, and inversions in 3D space. An E(3) equivariant mapping ensures the Hamiltonian matrix transforms properly under these symmetry operations. These inherent symmetries are captured with the proposed E(3) equivariant graph neural network in predicting the instantaneous Hamiltonian.
Unlike previous models that approximated equivariance through data augmentation, our model explicitly constructs the Hamiltonian matrix to strictly satisfy the inherent equivariance constraints of physical systems. The proposed model represents node and edge features using a direct sum of irreducible O(3) equivariant representations with different rotation orders. It updates these features through an equivariant message-passing function, then transforms them into on-site and off-site Hamiltonian matrix elements from the node and edge features, respectively. This equivariant construction of the Hamiltonian matrix allows $\mathrm{N^2AMD}$ to demonstrate excellent transferability and generalization, accurately predicting the electronic structure of large crystals outside its training set.

In the implementation of NAMD with the CPA, the movement of nuclei can be approximated by employing a precalculated AIMD trajectory. To generate this trajectory, a machine learning-based force field can be used. For a more precise comparison between $\mathrm{N^2AMD}$ and conventional DFT-NAMD, we utilize Allergo\cite{Musaelian_learning_2023} to produce trajectories for all NAMD simulations discussed here. It's noteworthy that Allergo also leverages an E(3) equivariant graph neural network to train the force field, which significantly improves accuracy in solid-state systems.

\subsection{\label{results23} Benchmark of $\mathrm{N^2AMD}$ }

Before demonstrating the capability of $\mathrm{N^2AMD}$, we first benchmark its performance on predicting both ground state properties and excited state dynamics. Rutile $\mathrm{TiO_2}$ and GaAs are chosen as the prototypical systems, while using $\mathrm{MoS_2}$ and Silicon to verify the generalizability of $\mathrm{N^2AMD}$. These materials have attracted extensive interest due to their promising applications in optoelectronics and solar energy\cite{ORegan_low-cost_1991, Linsebigler_Photocatalysis_1995, Nakata_tio2_2012}, and their carrier dynamics have been widely studied in recent years\cite{akimov_theoretical_2013, chu_ultrafast_2016, you_Correlated_2024}. Moreover, conventional DFT methods, using the PBE functional, notably underestimate the band gap of rutile $\mathrm{TiO_2}$ and GaAs.
Experimentally, these are observed to be 3.0 eV\cite{Amtout_optical_1995} and 1.4 eV\cite{Blakemore_Semiconducting_1982} respectively, in contrast to the DFT predictions of 1.88 eV\cite{landmann_electronic_2012} and 0.62 eV. Such discrepancies suggest that conventional DFT-NAMD may not align well with experimental results, highlighting the need for more accurate simulation methods like $\mathrm{N^2AMD}$.

\begin{figure}[t]
\centering
\includegraphics[width=1.0\linewidth]{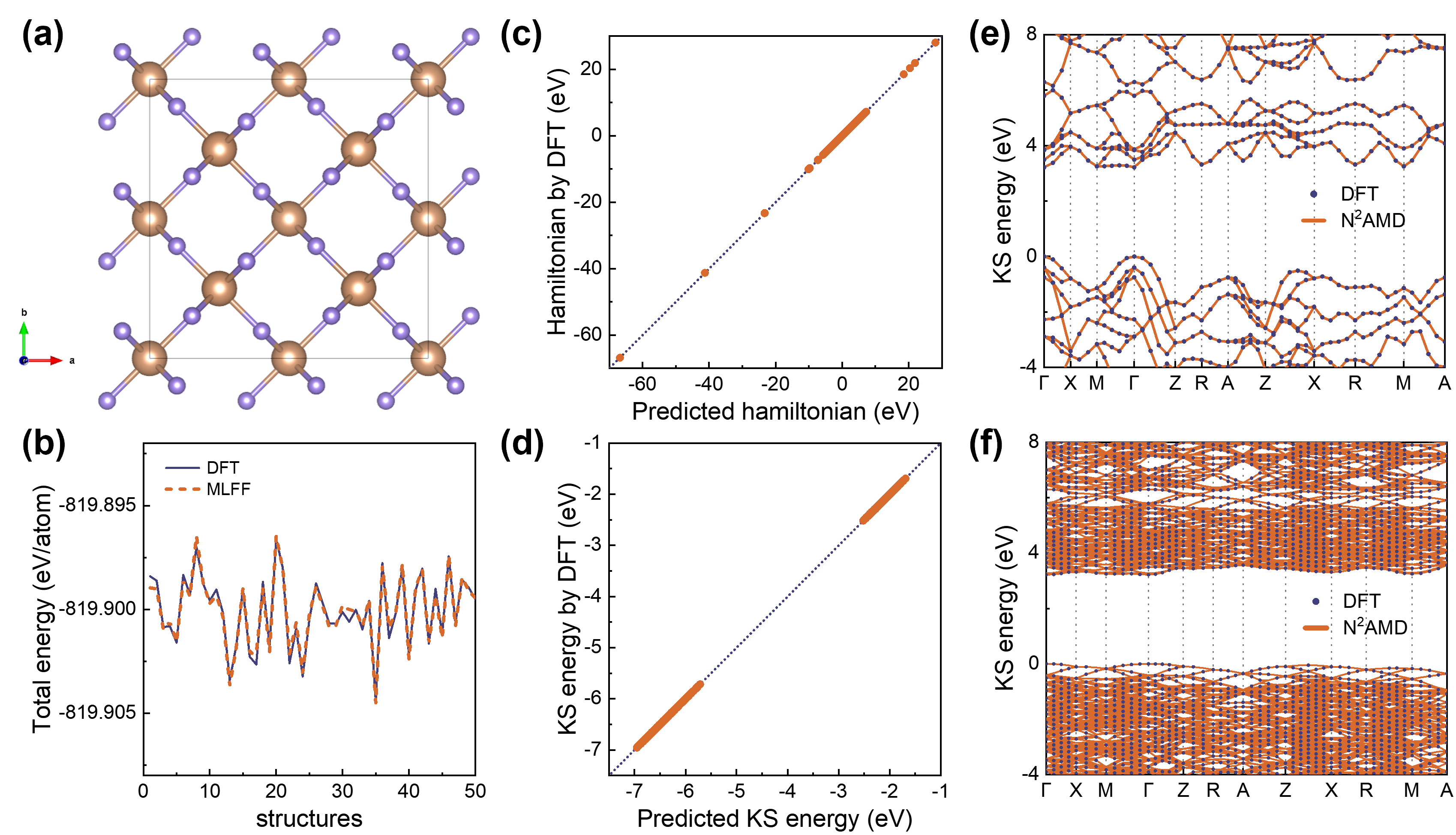}
\caption{\label{fig2} Benchmarking $\mathrm{N^2AMD}$ on ground state properties. (a) Geometry structure of stoichiometric rutile $\mathrm{TiO_2}$. (b) Comparison of MLFF predicted and DFT calculated potential energy on 50 random perturbed structures at 300K. (c-d) Comparison of $\mathrm{N^2AMD}$ predicted and DFT calculated Hamiltonian matrix elements and KS orbital energies on the test set. (e-f) Band structures of $1 \times 1 \times 1$ and $3 \times 3 \times 4$ supercells calculated by $\mathrm{N^2AMD}$ and DFT respectively.}

\end{figure}

We begin with benchmarking the machine learning force field (MLFF), which is essential for generating precalculated trajectories in NAMD simulations implemented with CPA. To validate the MLFF, we utilized 50 randomly perturbed structures. The potential energy profiles obtained from both the $\mathrm{N^2AMD}$ and DFT calculations are shown in Figure~\ref{fig2}b. Although the fluctuations in potential energy among the perturbed structures are relatively small, our ML model excellently reproduces the variations observed in DFT calculations.

In addition to the MLFF, an accurate description of the Hamiltonian and electronic structure for each snapshot along the trajectory is crucial for reliable NAMD simulations. As depicted in Figure~\ref{fig2}c and Figure S1a, the Hamiltonian matrix elements predicted by $\mathrm{N^2AMD}$ perfectly match the DFT-HSE06 results for both $\mathrm{TiO_2}$ and GaAs. By diagonalizing the Hamiltonian matrix, we simultaneously obtain the eigenvalues (Kohn-Sham orbital energies) and eigenvectors (Kohn-Sham wavefunctions). The fitting performance of the former is shown in Figure~\ref{fig2}d and Figure S1b for $\mathrm{TiO_2}$ and GaAs respectively, while the latter is presented in Figure S3 in the supplementary information\cite{si}.

We further present the band structures of $1 \times 1 \times 1$ and $3 \times 3 \times 4$ supercells of $\mathrm{TiO_2}$, calculated using $\mathrm{N^2AMD}$ and DFT-HSE06, in Figure~\ref{fig2}e and \ref{fig2}f, respectively. The mean average error (MAE) for the valence band maximum (VBM) and conduction band minimum (CBM) energy levels is consistently below 2.5 meV for both the primitive cell and supercells, even up to 216 atoms. Moreover, $\mathrm{N^2AMD}$ accurately reproduces the bandgap, band dispersion, and density of states, which are crucial for NAMD simulations. The predicted bandgap by $\mathrm{N^2AMD}$ is 3.219 eV for all simulation cells, remarkably close to the DFT-HSE06 calculation results of 3.212 eV ($1 \times 1 \times 1$ cell), 3.212 eV ($2 \times 2 \times 2$ cell), and 3.215 eV ($3 \times 3 \times 4$ cell). In comparison, the bandgap calculated by DFT-PBE is significantly underestimated at 1.788 eV.

Upon thoroughly examining the ground state properties predicted by $\mathrm{N^2AMD}$, we shifted our focus to benchmarking the excited properties. NACs are key quantities in NAMD but come with extremely high computational costs. Extensive efforts have been dedicated to developing ML models for NAC prediction. However, as indicated in eq~\ref{eq1c}, NAC depends on the derivative of the Hamiltonian and nuclear velocities, making NAC prediction considerably less accurate compared to FF prediction. Instead of directly predicting NAC, we opted to numerically compute NAC using the ML Hamiltonian. This approach ensures that the accuracy of the prediction of NAC is on par with the prediction of FF, bypassing the inherent challenges associated with accurately predicting NAC values.

\begin{figure}[t]
\centering
\includegraphics[width=1.0\linewidth]{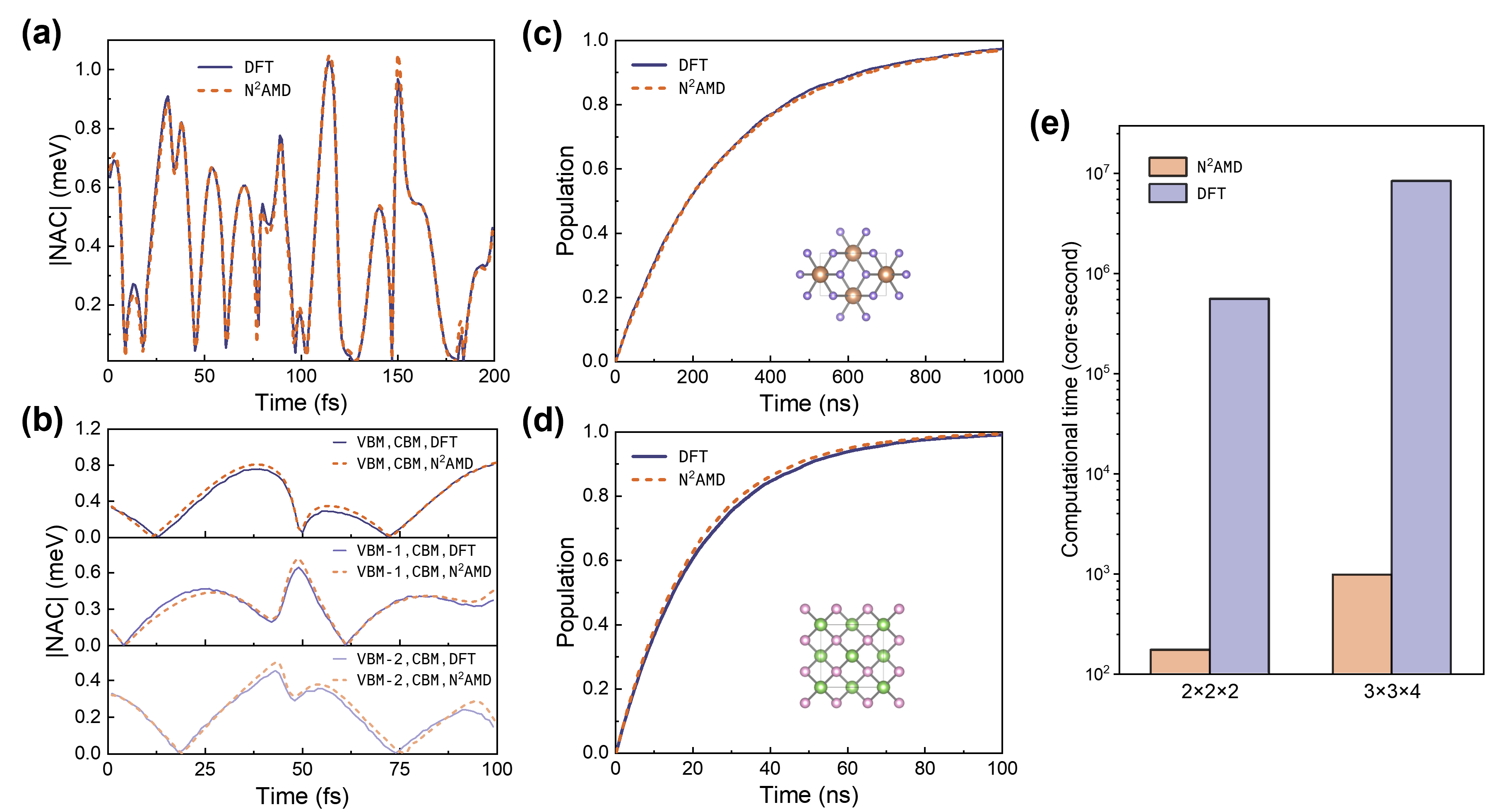}
\caption{\label{fig3} Benchmarking $\mathrm{N^2AMD}$ on excited state properties. (a-b) Absolute values of NACs between VBM and CBM along a short trajectory for (a) $\mathrm{TiO_2}$ and (b) GaAs, calculated by $\mathrm{N^2AMD}$ and DFT-HSE06 respectively. (c-d) The demonstrative non-radiative electron-hole recombination for $\mathrm{TiO_2}$ and GaAs, calculated by $\mathrm{N^2AMD}$ and DFT-HSE06 respectively. (e) Comparison of
computational resources for one single-step calculation required by two methods.}
\end{figure}

To validate the accuracy of the NACs predicted by $\mathrm{N^2AMD}$, we generated a 200 fs MD trajectory using MLFF and calculated the electronic structure at each timestep using both $\mathrm{N^2AMD}$ and DFT. Figure S2a-b depicts the Kohn-Sham orbital energies of VBM and CBM along the trajectory\cite{si}. The difference between VBM and CBM band energy, computed by $\mathrm{N^2AMD}$ and DFT method, is barely noticeable. Furthermore, the absolute NAC values between VBM and CBM (Figure~\ref{fig3}a-b) for both systems exhibit negligible differences. For GaAs, due to the three-fold degeneracy of the VBM, we present the NAC values of CBM \& VBM, VBM-1, and VBM-2. Notably, $\mathrm{N^2AMD}$ accurately predicts both peak and near-zero values of NACs, with MAE of only 0.017 meV and 0.035 meV for $\mathrm{TiO_2}$ and GaAs, respectively. After that, we employ the DISH approach to simulate the non-radiative recombination in both systems. Given the notorious computational cost of hybrid functional in ab initio calculation of comparison sets, we replicated this short 200 fs MD trajectory for the entire NAMD trajectory to evaluate the accuracy of $\mathrm{N^2AMD}$. Figure~\ref{fig3}c-d demonstrates a consistent evolution of carrier population and recombination rate throughout the dynamics between the ML and DFT methods. It is worth emphasizing that the obtained recombination rate presented here is less meaningful, as it serves as a benchmark test for the predictive capability of $\mathrm{N^2AMD}$ and lacks sufficient sampling. In reality, accurate results can only be obtained using the proposed ML model at the hybrid functional level, which will be investigated later.

The computational cost of $\mathrm{N^2AMD}$ is significantly lower than that of DFT-NAMD. Specifically, the computational expense of processing a single snapshot for $2\times2\times2$ or $3\times3\times4$ systems in $\mathrm{N^2AMD}$ is reduced by four orders of magnitude, as shown in Figure~\ref{fig3}e. It should be noted that HONPAS, which we used for comparison with our ML frameworks, employs the NAO2GTO scheme to compute electron repulsion integrals and their derivatives, making it substantially faster than other widely used DFT codes\cite{qin_honpas_2015}. However, even with this optimization, a NAMD simulation requires at least thousands of such single-point calculations, rendering the use of hybrid functionals in DFT-NAMD impractical due to the high computational cost.

\begin{figure}[t]
\centering
\includegraphics[width=1.0\linewidth]{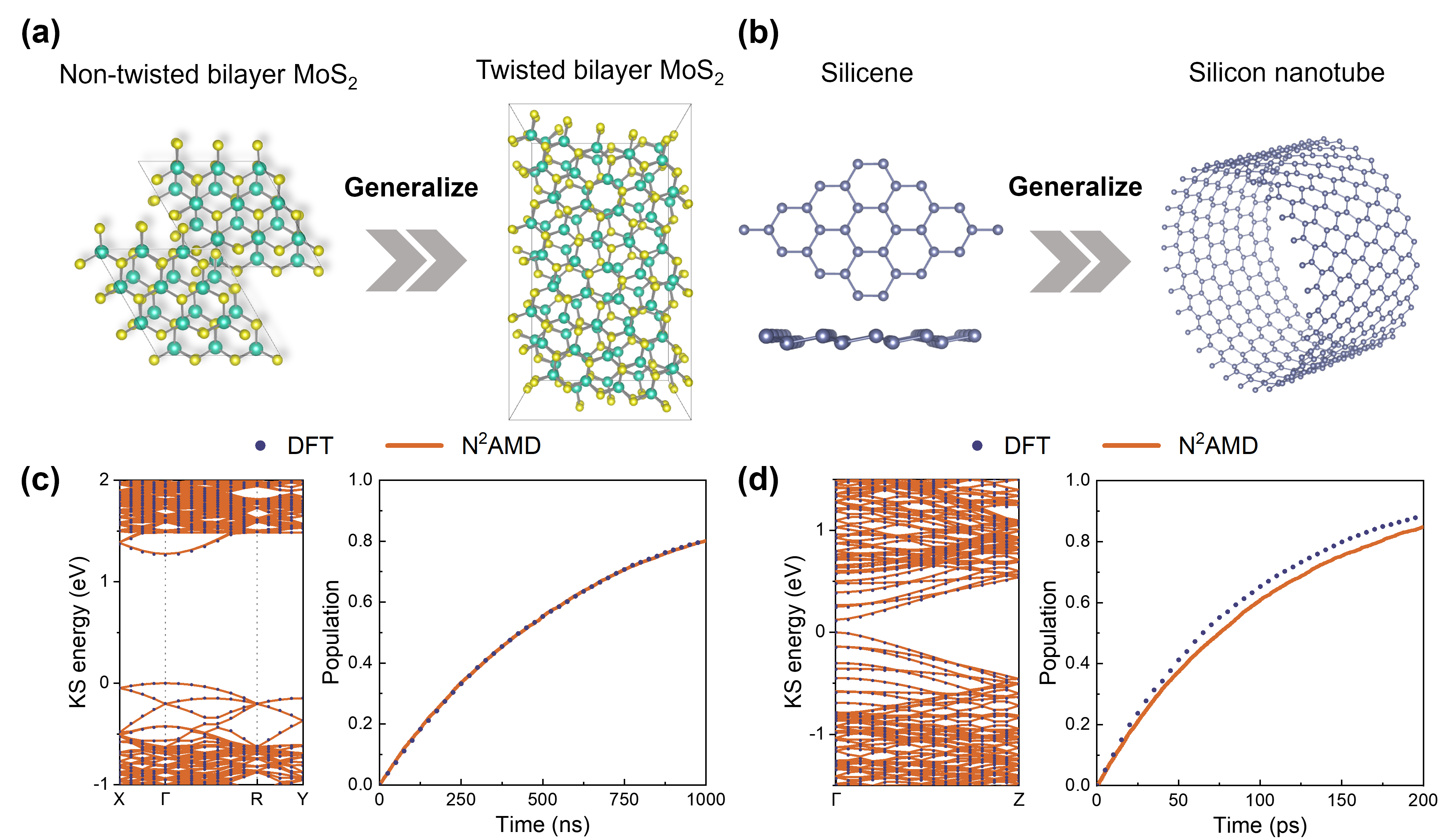}
\caption{\label{fig7} Generalization capability of $\mathrm{N^2AMD}$. (a) and (c) the use of $\mathrm{N^2AMD}$, trained by DFT results of non-twisted bilayer $\mathrm{MoS_2}$, to predict the band structure and carrier dynamics in twisted bilayer $\mathrm{MoS_2}$. (b) and (d) the use of $\mathrm{N^2AMD}$, trained by DFT results of monolayer Silicon, to predict the band structure and carrier dynamics in the Silicon nanotube. }
\end{figure}

We further evaluate the generalization capability of $\mathrm{N^2AMD}$ by predicting properties of new structures notably different from those in the training set, as shown in Figure~\ref{fig7}a-b. Given that employing hybrid functional in DFT-NAMD is infeasible for those large structures, we employed the PBE functional in both simulations to ensure consistent comparison. In the first case, we trained the model on the non-twisted bilayer $\mathrm{MoS_2}$ and then used it to predict the band structure and perform NAMD simulations on a twisted bilayer. In the second case, the model was trained using silicene and subsequently applied to simulate a curved nanotube. As shown in Figure~\ref{fig7}c-d and Figure S4\cite{si}, $\mathrm{N^2AMD}$ successfully reproduces both the band structure, NACs, and real-time dynamics for each case. Note that both twisted materials and nanotubes feature significantly large cells, making DFT calculation costly. In contrast, $\mathrm{N^2AMD}$ offers accurate simulations at a significantly reduced computational cost.

\subsection{\label{resultsB} Application to hybrid functional NAMD }

Following the benchmarks, we investigate the non-radiative electron-hole recombination dynamics in stoichiometric $\mathrm{TiO_2}$ and GaAs completely by $\mathrm{N^2AMD}$. In addition to hybrid functional calculations, we conducted simulations with PBE functional to establish a comparative analysis.

In NAMD simulations, the e-h recombination timescales predominantly depend on the band gap, pure-dephasing time, and NAC between donor and acceptor states. Typically, a larger band gap, shorter pure-dephasing time, and weaker NAC contribute to slower e-h recombination rates.

\begin{figure}[t]
\centering
\includegraphics[width=1.0\linewidth]{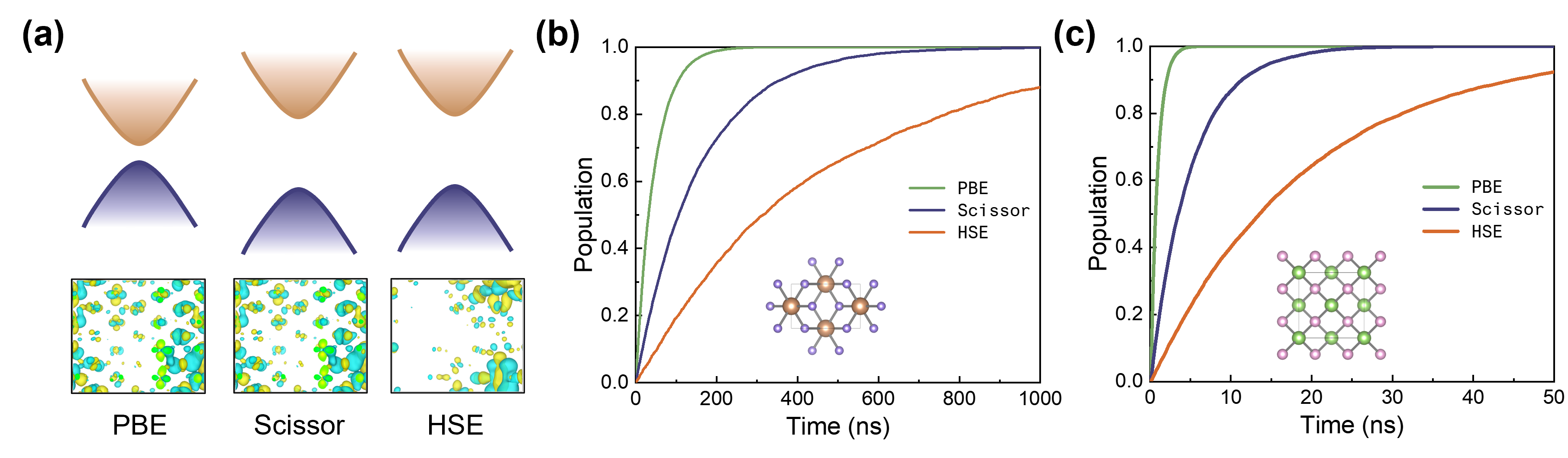}
\caption{\label{fig4} Nonradiative Recombination dynamics predicted by different exchange-correlation functionals. 
(a) Schematic band structures and frontier orbital wavefunctions calculated by PBE, scissor-corrected PBE, and HSE.
(b-c) Different exchange-correlation functionals predicted (b) the evolution of hole population on CBM in $\mathrm{TiO_2}$ and (c) the evolution of electron population on VBM in GaAs.
}
\end{figure}
\begin{table}[t]
\caption{\label{tab1}The fundamental bandgaps, pure dephasing times, (canonically averaged root mean squared) NACs, and e-h nonradiative lifetimes of stoichiometric $\mathrm{TiO_2}$ and GaAs calculated by different exchange-correlation functionals.}
\renewcommand{\arraystretch}{1.2}
\begin{tabular*}{0.85\textwidth}{@{\extracolsep\fill}clccc}
\toprule
& Functional & DFT-PBE & DFT-Scissor & ML-HSE06 \\
\midrule
\multirow{4}{*}{\quad$\mathrm{TiO_2}$} & Bandgap (eV)   & 1.79 & 3.22 & 3.22   \\
                                       & Dephasing (fs) & 16.5 & 16.5 & 15.3   \\
                                       & NAC (meV)      & 0.77 & 0.77 & 0.43   \\
                                       & Lifetime (ns)  & 45.6 & 153.7 & 465.6 \\
\midrule
\multirow{5}{*}{\quad GaAs} & Bandgap (eV)                                  & 0.61 & 1.39 & 1.39 \\
                            & NAC $d_{\mathrm{VBM-2}}^{\mathrm{CBM}}$ (meV) & 0.86 & 0.86 & 0.41 \\ 
                            & NAC $d_{\mathrm{VBM-1}}^{\mathrm{CBM}}$ (meV) & 0.81 & 0.81 & 0.53 \\
                            & NAC $d_{\mathrm{VBM}}^{\mathrm{CBM}}$ (meV)   & 0.61 & 0.61 & 0.43 \\
                            & Lifetime (ns)                                 & 0.84 & 4.9 & 19.4\\
\botrule
\end{tabular*}
\end{table}

It is well-known that the PBE functional significantly underestimates the band gap compared to hybrid functionals\cite{heyd_hybrid_2003}. For $\mathrm{TiO_2}$, the band gap is underestimated by 1.43 eV, and for GaAs, by 0.78 eV (Table~\ref{tab1}). To rectify the band gap underestimation in NAMD simulations employing the PBE functional, a scissor operation is frequently utilized. Although the scissor correction applied in NAMD can adjust the band gap, it does not alter the wavefunctions, as depicted in Figure~\ref{fig4}a. Consequently, as shown in Table~\ref{tab1}, the canonically averaged root mean square values of NACs, calculated using PBE and Scissor correction, are significantly overestimated for both $\mathrm{TiO_2}$ and GaAs. Given these substantial deviations in band gap and NAC values, the electron-hole recombination time calculated using the HSE06 functional is approximately 10 to 20 times longer than that computed with the PBE functional (Table~\ref{tab1}, Figure~\ref{fig4}b-c). Moreover, as indicated in Table~\ref{tab1}, even after applying a scissor correction, the calculated lifetimes are still underestimated by approximately a factor of 3 to 4. This discrepancy can be attributed to the fact that while the scissor operation corrects the bandgap, it leaves NAC unchanged. Since NAC is dependent on both the band gap and the wavefunctions, adjustments solely to the band gap are insufficient for achieving accurate simulation outcomes. Therefore, $\mathrm{N^2AMD}$ provides a more rigorous and self-consistent approach for performing NAMD simulations in nanoscale and condensed matter systems.

\subsection{\label{results24} Application to large-scale NAMD simulation}

NAMD simulations are typically conducted using small simulation cells to manage the high computational costs associated with larger simulation cells. However, the use of small simulation cells often suffers from the finite size effect and leads to a significantly higher effective carrier density compared to realistic conditions. Consequently, the calculated lifetimes of non-equilibrium charge carriers often differ from experimental results by orders of magnitude. Moreover, to capture emergent properties such as anharmonicity, geometry reconstruction, symmetry breaking, and disorder, large-scale NAMD simulations are crucial. These properties are essential for a more accurate depiction of carrier dynamics. The $\mathrm{N^2AMD}$ framework addresses this by significantly reducing the computational burden, making it feasible to simulate complex systems on a scale several orders of magnitude larger than what is possible with conventional DFT-based methods.

\begin{figure}[t]
\centering
\includegraphics[width=1.0\linewidth]{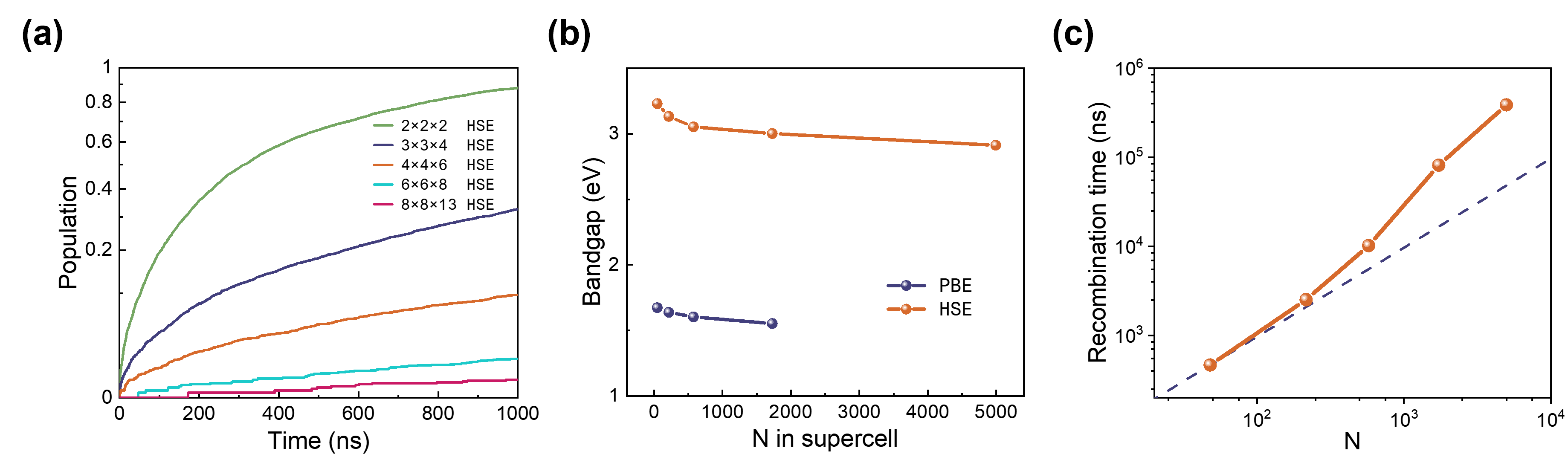}
\caption{\label{fig5} Large-scale hybrid functional NAMD simulation of recombination dynamics in stoichiometric $\mathrm{TiO_2}$. 
(a) The evolution of hole population on CBM, computed by different supercells from $2\times2\times2$ with 48 atoms to $8\times8\times13$ with 4992 atoms. 
(b) PBE and HSE predicted canonically averaged bandgap at 300K as the number of atoms in the supercell increases from 48, 216, 576, 1728 to 4992.
(c) The scaling relationship between recombination lifetime and supercell size. the dashed line indicates a proportional relation while the dotted line is the results in this work.
}
\end{figure}

Here, we explored carrier dynamics of $\mathrm{TiO_2}$ across various simulation cell sizes ranging from $2\times2\times2$ (48 atoms) to $8\times8\times13$ (4992 atoms), using the hybrid functional. The pure dephasing time, NACs, and calculated lifetime for various simulation cell sizes are tabulated in Table~\ref{tab2}. We observed that the predicted band gap at 0 K for all simulation sizes is consistent at 3.219 eV. However, at 300 K, the canonically averaged band gap decreases with increases in the cell size. This trend was confirmed through DFT calculations using the PBE functional for cells ranging from $2\times2\times2$ to $6\times6\times8$, and a consistent decrease in the averaged band gap with increasing cell size was noted (Figure~\ref{fig5}b). Further analysis of the wavefunctions of frontier orbitals revealed that this reduction in band gap is due to the localization of CBM, as shown in Figure S5b and S5d\cite{si}. Such a peculiar localized state has been overlooked for a long time and merits further thorough investigation. Regarding the pure dephasing time between VBM and CBM, which measures the coherence between these states and is another critical parameter in NAMD simulations, it was found to be 16.5 fs for PBE and 15.3 fs for HSE06 with a $2\times2\times2$ supercell. These results suggest that the coherence between these states is relatively insensitive to the choice of functional in this system.

\begin{table}[t]
\caption{\label{tab2}The canonically averaged bandgaps, pure dephasing times, (root mean squared) NACs and e-h recombination lifetime of stoichiometric $\mathrm{TiO_2}$ calculated by different supercell sizes.}
\begin{tabular*}{\textwidth}{@{\extracolsep\fill}lcccc}
\toprule
\quad Supercell & Averaged Bandgap (eV) & Dephasing (fs) & NAC (meV) 
& Lifetime (ns)\\
\midrule
\quad $2\times2\times2$   & 3.23 & 15.3 & 0.43 & $4.66\times10^2$ \\ 
\quad $3\times3\times4$   & 3.13 & 18.9 & 0.19 & $2.52\times10^3$ \\ 
\quad $4\times4\times6$   & 3.05 & 17.5 & 0.099 & $1.02\times10^4$ \\ 
\quad $6\times6\times8$   & 3.00 & 16.0 & 0.043 & $8.17\times10^4$ \\ 
\quad $8\times8\times13$  & 2.91 & 18.2 & 0.022 & $3.89\times10^5$ \\ 
\botrule
\end{tabular*}
\end{table}

Previous research\cite{chu_low-frequency_2020, wang_Effective_2022, Shakiba_Dependence_2023} suggested that effective carrier lifetimes in pristine semiconductors under realistic conditions scale inversely with carrier density. As the size of the simulation cell increases, we observe a corresponding decrease in NACs, which leads to extended carrier lifetimes. Figure~\ref{fig5}c shows that, for a simulation cell size of $3\times3\times4$, the lifetime remains proportional to the number of atoms in the supercell. However, when employing even larger simulation cells, the lifetime is significantly prolonged. This phenomenon can be attributed to the formation of the localized state that effectively suppresses the NAC and thus extends the carrier lifetime.

Our findings highlight the necessity of using large-scale simulations to accurately capture the properties and behaviors of carriers in nanoscale and condensed matter systems. The $\mathrm{N^2AMD}$ framework facilitates these simulations by drastically reducing the computational costs involved, making it feasible to explore emergent properties and uncover unique phenomena with large-scale simulation. These efforts are crucial for advancing our understanding of material behaviors that conventional DFT-based methods have previously missed.

\subsection{\label{results25} Application to defect-associated carrier dynamics}

NAMD simulations play a crucial role in the design of semiconductor devices such as solar cells, LEDs, and transistors, where effectively controlling and understanding recombination processes is vital. In these materials, defects can either capture or scatter charge carriers, impacting essential properties such as conductivity and luminescence. It is particularly important to comprehend how these defects influence non-radiative decay processes. NAMD offers valuable insights into the dynamics of carriers associated with defects. However, when DFT-based NAMD employs conventional exchange-correlation functionals such as LDA or GGA, it frequently misestimates certain properties of defects. Such inaccuracies can lead to skewed predictions regarding defect energies, charge transition levels, and defect formation energies, resulting in discrepancies between experimental results and theoretical predictions. Recently, hybrid functionals are increasingly acknowledged for their pivotal role in more accurately depicting the dynamics of carriers associated with defects\cite{janotti_hybrid_2010, zhang_Origin_2022}.

\begin{figure}[t]
\centering
\includegraphics[width=0.7\linewidth]{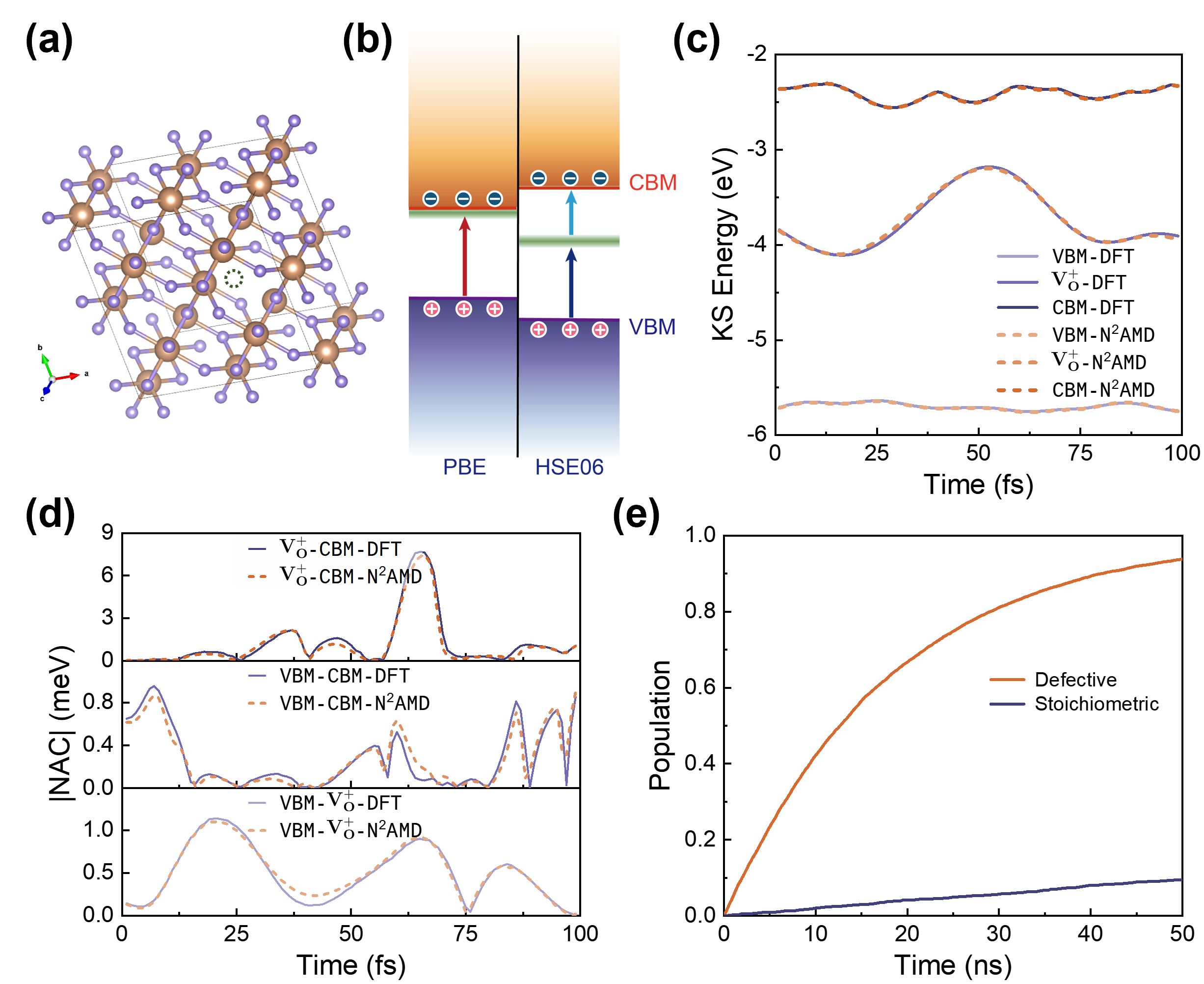}
\caption{\label{fig6}NAMD simulation of $V_O^+$ defect-associated recombination dynamics. (a) Geometry structure of $V_O^+$ rutile $\mathrm{TiO_2}$. (b) Schematic $V_O^+$ defect energy level alignment (green line) and e-h recombination path using PBE and HSE06 functionals. (c) Benchmark of KS orbital energies of VBM, defect state, and CBM, (d) benchmark of absolute values of NACs between VBM, defect state, and CBM along a 100 fs MD trajectory. Only the spin-up channel is plotted. (e) Comparison of the evolution of hole population on CBM with and without defect assistance, calculated by $\mathrm{N^2AMD}$.}
\end{figure}

Taking the example of the positively charged oxygen vacancy ($V_O^+$) in $\mathrm{TiO_2}$ (Figure~\ref{fig6}a), this defect has been experimentally confirmed to act as an electron-hole recombination center\cite{maity_study_2018}. However, as shown in Figure~\ref{fig6}b, the widely used PBE functional typically predicts that it is a shallow level near the CBM, suggesting it has minimal impact on electron-hole recombination. The accurate defect level is only predictable when employing a hybrid functional, highlighting the critical importance of using these advanced functionals to precisely describe deep defect levels and their recombination dynamics. The bandstructure calculated with DFT employing PBE and HSE06 functional are shown in Figure S6\cite{si}. 
In addition to the need for high-level functionals, investigating such defects requires the use of spin-polarized calculations with an additional positive charge, which offers a playground for $\mathrm{N^2AMD}$ to demonstrate its capability in more complex systems. By leveraging the accuracy and transferability of our model, we can explore defect-assisted recombination processes using hybrid functionals. We also perform DFT calculations for a short trajectory to validate the effectiveness of $\mathrm{N^2AMD}$ for this complex system. Figure~\ref{fig6}c and \ref{fig6}d demonstrate the excellent fitting performance of $\mathrm{N^2AMD}$ on both KS orbital energies and NACs. Employing $\mathrm{N^2AMD}$, the lifetime of the defect state is calculated to be 18.1 ns, significantly shorter than that in a stoichiometric system, as depicted in Figure~\ref{fig6}e. This finding aligns with experimental observations\cite{Furube_charge_1999, yamada_determination_2012}, highlighting the substantial role played by the oxygen vacancy as an electron-hole recombination center. 
It should be noted that these insights are only accessible through our proposed ML model. These findings emphasize the importance of using high-level functionals and ML models to bridge the gap between experimental observations and theoretical predictions in the study of defect-related phenomena in materials. 

\section{Discussion}\label{sec12}

The incorporation of E(3) equivariance as prior knowledge into the message-passing deep-learning framework for Hamiltonians has significantly enhanced both efficiency and accuracy. A key advantage of $\mathrm{N^2AMD}$ is its ability to learn from DFT Hamiltonians that employ high-level exchange-correlation functionals on small structures, and then accurately predict Hamiltonians for various structures without additional DFT calculations. Essential quantities for NAMD, such as eigenvalues and NACs, are computed directly from the predicted Hamiltonian. This approach ensures efficiency and accuracy in evaluating these quantities, particularly for those that depend on Hamiltonian derivatives. Furthermore, because $\mathrm{N^2AMD}$ is designed to predict Hamiltonians, all other necessary quantities for advanced NAMD simulations can be readily acquired with $\mathrm{N^2AMD}$.

NAMD simulations, particularly in condensed matter materials, have seen significant advancements when combined with CPA. CPA streamlines NAMD simulations by omitting the back-reaction of electronic transitions on nuclear motions, allowing for the use of precomputed MD trajectories. This approach significantly reduces computational demands, making techniques such as surface hopping more practical for various applications in nanoscale and condensed matter systems. CPA is most effective in systems where atomic motions are driven by finite temperature rather than by electronic excitations. However, CPA's inability to account for real-time excited state forces limits its utility in simulating light-matter interactions, chemical reactions, and phase transitions.

One of the major challenges in moving beyond CPA-NAMD involves the calculation of NACVs. NACVs are essential because they quantify how much the electronic wavefunctions of the two states overlap and change as a function of nuclear positions. Accurately calculating NACVs is a computationally demanding task that requires the precise determination of electronic wavefunctions and their gradients with respect to nuclear coordinates. In NAMD simulations, the need for on-the-fly computation of NACVs at every timestep, typically using Density Functional Perturbation Theory, poses a formidable challenge, especially in nanoscale and condensed matter systems. The proposed workflow aims to revolutionize this aspect of NAMD simulations by integrating a GNN, which can transform the traditionally expensive electronic structure calculations. By doing so, it makes evaluating NACVs on the fly feasible\cite{si}. This innovation could potentially transform the landscape of NAMD simulations, facilitating more precise and dynamic modeling of non-adiabatic processes in 
nanoscale and condensed matter systems.

The proposed workflow is designed to transform the electronic structure in NAMD calculations by incorporating GNN. This approach effectively leverages an ML-based Hamiltonian to compute all essential quantities in NAMD, while keeping the core NAMD framework intact. This seamless integration ensures that it can be easily adopted in conjunction with other state-of-the-art NAMD methodologies that have been developed recently. For instance, Zhao and co-workers\cite{zheng_ab_2023} have introduced a momentum-space NAMD algorithm that can model the relaxation dynamics of electrons by substituting NAC with electron-phonon couplings (EPC). However, the substantial computational demands of calculating EPC limit this algorithm to small-scale systems. Our proposed workflow can overcome this challenge by efficiently calculating EPC and other necessary quantities using the ML-Hamiltonian\cite{zhong_Accelerating_2023}. This enhancement could significantly broaden the applicability of Zhao's method to more complex situations, such as twisted materials or systems with defects.

In addition to momentum-space NAMD, spin dynamics is another area attracting significant interest\cite{zheng_Spin-orbit_2022}, particularly for examining carrier dynamics in topological insulators and valley dynamics in novel 2D materials. Incorporating spin-orbit coupling (SOC) into NAMD simulations is essential for these studies. The proposed framework can seamlessly integrate both the adiabatic and diabatic representations of surface hopping with SOC\cite{zhong_Accelerating_2023-1}. In the adiabatic representation, our framework predicts the Hamiltonian with non-collinear SOC and diagonalizes it to obtain the spinor-based wavefunctions. Conversely, in the diabatic representation, where SOC is treated as a perturbation to the collinear Hamiltonian, we can predict the collinear Hamiltonian and directly evaluate the SOC Hamiltonian. This dual capability allows for comprehensive and versatile modeling and understanding of the spin dynamics.

To conclude, we present $\mathrm{N^2AMD}$, an innovative NAMD workflow enhanced by E(3)-equivariant ML models. This approach enables efficient and accurate NAMD simulations of large-scale materials at the level of hybrid-functional accuracy. We demonstrate several cases where conventional NAMD approaches fail due to limitations in computational efficiency and accuracy. We envision that this framework can be effectively combined with the latest advancements in NAMD technology, potentially leading to transformative developments in NAMD methodology and advancing the field of physical science research. By addressing the limitations of existing methods and expanding their applicability to complex systems and high-level theory, $\mathrm{N^2AMD}$ opens up new possibilities for investigating and understanding the dynamic properties of materials at an unprecedented scale and accuracy.

\section{\label{theory}Methods}

\subsection{Computational Details of DFT calculations}
DFT calculations of $\mathrm{TiO_2}$ and GaAs are performed with the HONPAS code\cite{qin_honpas_2015}, which implements NAO basis and norm-conserving pseudopotentials (NCPP). 
The valence electron configuration is $3s^2 3p^6 3d^2 4s^2$ for Ti atoms, $2s^2 2p^4$ for O atoms, $4s^2 4p^1$ for Ga atoms and $4s^2 4p^3$ for As atoms. Therefore Ti-3s2p2d, O-2s2p1d, Ga-2s2p1d, and As-2s2p1d NAOs are applied to expand the Hamiltonian matrix and wavefunctions.
The Heyd-Scuseria-Ernzerhof (HSE06) hybrid functional\cite{ heyd_hybrid_2003} employing a mixing parameter $\alpha=0.25$ and a range separation parameter $\omega=0.11 
 \mathrm{Bohr^{-1}}$, is used in above calculations.
DFT calculations of $\mathrm{MoS_2}$ bilayers, silicenes, and silicon nanotubes are performed in the PBE functional via the OpenMX software\cite{ozaki_Variationally_2003}. The van der Waals corrections are considered by using the DFT-D3 method. Mo-3s2p2d, 
S-2s2p1d and Si-2s2p1d NAOs are employed in the OpenMX simulations. 

The experiment result\cite{GRANT_properties_1959} for the lattice parameters of $\mathrm{TiO_2}$ is adopted and kept fixed in this work. 
The unit cell lattice constants are $4.59, 4.59, 2.956 
 \si{\angstrom}$, all perpendicular to each other. A $6 \times 6 \times 10$ Gamma-centered k-mesh is used to sample the Brillouin zone. 
To simulate the $V_O^+$ defect in $\mathrm{TiO_2}$. A $2 \times 2 \times 2$ supercell is utilized. One electron is subtracted and one oxygen atom, indicated by a dashed circle, is removed as shown in Figure~\ref{fig6}a. Spin polarization is employed for the calculation of the defective system, and the spin momentum is fixed to the up channel to maintain continuity over time.
For GaAs, a 5.6537\si{\angstrom} cubic lattice, a $2 \times 2 \times 1$ supercell, and a $4 \times 4 \times 8$ k-grid are utilized in the calculations.
The twisted $\mathrm{MoS_2}$ bilayer, with 42 atoms in its primitive cell, is composed of stacked supercells with different spatial orientations. The twist angle of it is $38.2^{\circ}$. To fold its CBM to the $\Gamma$ point for recombination calculations, we further expand it to a $3 \times \sqrt{3}$ orthogonal supercell with 252 atoms. 
The zigzag (30,0) silicon nanotube is generated by fully relaxed silicene. The diameter of the nanotube is 36.9$\si{\angstrom}$. The thickness of the vacuum layer is $25\si{\angstrom}$ for 2D materials and $15\si{\angstrom}$ for the nanotube.
Only $\Gamma$ point is used in our k-mesh for twist-angle $\mathrm{MoS_2}$ bilayer and silicon nanotube calculations.

\subsection{Computational Details of NAMD calculations}

The e-h recombination dynamics is performed via Hefei-NAMD code\cite{zheng_ab_2019}, which implements the ab initio NAMD algorithm under CPA.
The DISH algorithm is adopted to account for the decoherence effect in recombination. 
Phase correction is used to correct the random phase of the Bloch wavefunctions in the adiabatic representation\cite{akimov_simple_2018}.
For the benchmark of stoichiometric $\mathrm{TiO_2}$, a 200 fs microcanonical MD trajectory at 300 K is generated with a timestep of 1 fs using MLFF. The electronic structure of each atomic geometry on the trajectory is calculated by both $\mathrm{N^2AMD}$ and $\mathrm{HONPAS}$ to verify the effectiveness of our method.
For the complete NAMD simulation of both stoichiometric and defective $\mathrm{TiO_2}$, a 5000 fs microcanonical MD trajectory is produced fully by $\mathrm{N^2AMD}$. 
For GaAs, twist-angle $\mathrm{MoS_2}$ bilayer and silicon nanotube, a 1000 fs microcanonical MD trajectory is utilized. 
For the silicon nanotube, the MD simulation is performed at 50 K due to its instability under high temperatures.
For all materials studied, we conduct NAMD simulations using 20 different initial configurations. Each configuration is sampled with 200 trajectories.
To simulate the long-time dynamics of e-h recombination, the MD trajectory is concatenated head-to-tail as suggested by previous work\cite{zheng_ab_2019, prezhdo_modeling_2021}. The final NAMD results for all systems are obtained by averaging the results of all initial configurations and trajectories.

\subsection{Details of neural network training}
The datasets utilized to train the ML models consist of 1000 stoichiometric $\mathrm{TiO_2}$, 500 $V_O^+$ defective $\mathrm{TiO_2}$, 300 GaAs, 1000 non-twisted $\mathrm{MoS_2}$ bilayers, and 2000 silicenes, respectively. These structures are randomly selected from molecular dynamics (MD) trajectories, with temperatures ranging from 200 K to 500 K, except for silicene, which is heated from 5 K to 300 K. To improve the models' transferability, the dataset includes $\mathrm{MoS_2}$ bilayers varied by slide vectors and layer intervals, and silicenes subjected to stresses ranging from 0\% to 3\%. The model for each material is trained independently.

For the machine learning force field, Allegro\cite{Musaelian_learning_2023} with two layers, a max angular quantum number $l_{max}=2$, and a radius cutoff of $6 \si{\angstrom}$, is used.
The dataset is randomly divided into two subsets with a 9:1 ratio for training and validation. 
The initial learning rate is set to 0.002, and then reduced according to an on-plateau scheduler with a patience of 10 and a decay factor of 0.5.
The model is trained with a joint mean square error loss function that targets both per-atom energies and forces, and it is optimized using the Adam optimizer.
The training is finalized when the learning rate is dropped to $10^{-5}$.

For the Hamiltonian model, HamGNN\cite{zhong_transferable_2023} with five interaction layers and a radius cutoff of 20 Bohr is utilized.
The dataset is randomly split into training, validation, and test sets with a ratio of 7:2:1.
The training hyperparameters in HamGNN are similar to those in Allegro with a few exceptions: the initial learning rate is set at 0.001, the AdamW optimizer is employed, and the final learning rate is adjusted to $10^{-6}$. 
A two-stage training process is utilized in the training process.
In the first stage, the MAE of real-space Hamiltonian matrices is used as the loss function.
In the second stage, an extra regularization term representing the band energy error
is added to the loss function with a weight of 0.01 to improve the transferability and stability of the predictions.

\section*{Data and Code availability}
Data and code that support the results of this work will be made publicly available when the paper is formally accepted.

\section*{Acknowledgments}

We acknowledge the support of National Natural Science Foundation of China (11991061, X.G.G., 12274081, W.B.C., 12188101, H.J.X.); Shanghai Pilot Program for Basic Research - FuDan University 21TQ1400100 (22TQ017, W.B.C). O.V.P. acknowledges the support of the US National Science Foundation, grant CHE-2154367


\begin{thebibliography}{10}
\expandafter\ifx\csname url\endcsname\relax
  \def\url#1{\burl{#1}}\fi
\expandafter\ifx\csname urlprefix\endcsname\relax\def\urlprefix{URL }\fi
\providecommand{\bibinfo}[2]{#2}
\providecommand{\eprint}[2][]{\url{#2}}
\providecommand{\doi}[1]{\url{https://doi.org/#1}}
\bibcommenthead

\bibitem{zheng_ab_2023}
\bibinfo{author}{Zheng, Z.} \emph{et~al.}
\newblock \bibinfo{title}{Ab initio real-time quantum dynamics of charge carriers in momentum space}.
\newblock \emph{\bibinfo{journal}{Nature Computational Science}} \textbf{\bibinfo{volume}{3}}, \bibinfo{pages}{532--541} (\bibinfo{year}{2023}).

\bibitem{zheng_Multiple_2023}
\bibinfo{author}{Zheng, F.} \& \bibinfo{author}{Wang, L.-w.}
\newblock \bibinfo{title}{Multiple k -{Point} {Nonadiabatic} {Molecular} {Dynamics} for {Ultrafast} {Excitations} in {Periodic} {Systems}: {The} {Example} of {Photoexcited} {Silicon}}.
\newblock \emph{\bibinfo{journal}{Physical Review Letters}} \textbf{\bibinfo{volume}{131}}, \bibinfo{pages}{156302} (\bibinfo{year}{2023}).

\bibitem{nelson_non-adiabatic_2020}
\bibinfo{author}{Nelson, T.~R.} \emph{et~al.}
\newblock \bibinfo{title}{Non-adiabatic {Excited}-{State} {Molecular} {Dynamics}: {Theory} and {Applications} for {Modeling} {Photophysics} in {Extended} {Molecular} {Materials}}.
\newblock \emph{\bibinfo{journal}{Chemical Reviews}} \textbf{\bibinfo{volume}{120}}, \bibinfo{pages}{2215--2287} (\bibinfo{year}{2020}).

\bibitem{zheng_ab_2019}
\bibinfo{author}{Zheng, Q.} \emph{et~al.}
\newblock \bibinfo{title}{Ab initio nonadiabatic molecular dynamics investigations on the excited carriers in condensed matter systems}.
\newblock \emph{\bibinfo{journal}{WIREs Computational Molecular Science}} \textbf{\bibinfo{volume}{9}}, \bibinfo{pages}{e1411} (\bibinfo{year}{2019}).

\bibitem{long_Nonadiabatic_2017}
\bibinfo{author}{Long, R.}, \bibinfo{author}{Prezhdo, O.~V.} \& \bibinfo{author}{Fang, W.}
\newblock \bibinfo{title}{Nonadiabatic charge dynamics in novel solar cell materials}.
\newblock \emph{\bibinfo{journal}{WIREs Computational Molecular Science}} \textbf{\bibinfo{volume}{7}}, \bibinfo{pages}{e1305} (\bibinfo{year}{2017}).

\bibitem{zheng_phonon-assisted_2017}
\bibinfo{author}{Zheng, Q.} \emph{et~al.}
\newblock \bibinfo{title}{Phonon-{Assisted} {Ultrafast} {Charge} {Transfer} at van der {Waals} {Heterostructure} {Interface}}.
\newblock \emph{\bibinfo{journal}{Nano Letters}} \textbf{\bibinfo{volume}{17}}, \bibinfo{pages}{6435--6442} (\bibinfo{year}{2017}).

\bibitem{lian_ultrafast_2020}
\bibinfo{author}{Lian, C.}, \bibinfo{author}{Zhang, S.-J.}, \bibinfo{author}{Hu, S.-Q.}, \bibinfo{author}{Guan, M.-X.} \& \bibinfo{author}{Meng, S.}
\newblock \bibinfo{title}{Ultrafast charge ordering by self-amplified exciton–phonon dynamics in {TiSe2}}.
\newblock \emph{\bibinfo{journal}{Nature Communications}} \textbf{\bibinfo{volume}{11}}, \bibinfo{pages}{43} (\bibinfo{year}{2020}).

\bibitem{chu_ultrafast_2022}
\bibinfo{author}{Chu, W.} \emph{et~al.}
\newblock \bibinfo{title}{Ultrafast charge transfer coupled to quantum proton motion at molecule/metal oxide interface}.
\newblock \emph{\bibinfo{journal}{Science Advances}} \textbf{\bibinfo{volume}{8}}, \bibinfo{pages}{eabo2675} (\bibinfo{year}{2022}).

\bibitem{prezhdo_modeling_2021}
\bibinfo{author}{Prezhdo, O.~V.}
\newblock \bibinfo{title}{Modeling {Non}-adiabatic {Dynamics} in {Nanoscale} and {Condensed} {Matter} {Systems}}.
\newblock \emph{\bibinfo{journal}{Accounts of Chemical Research}} \textbf{\bibinfo{volume}{54}}, \bibinfo{pages}{4239--4249} (\bibinfo{year}{2021}).

\bibitem{lin_dependence_2016}
\bibinfo{author}{Lin, Y.} \& \bibinfo{author}{Akimov, A.~V.}
\newblock \bibinfo{title}{Dependence of {Nonadiabatic} {Couplings} with {Kohn}–{Sham} {Orbitals} on the {Choice} of {Density} {Functional}: {Pure} vs {Hybrid}}.
\newblock \emph{\bibinfo{journal}{The Journal of Physical Chemistry A}} \textbf{\bibinfo{volume}{120}}, \bibinfo{pages}{9028--9041} (\bibinfo{year}{2016}).

\bibitem{zhu_density_2021}
\bibinfo{author}{Zhu, Y.} \& \bibinfo{author}{Long, R.}
\newblock \bibinfo{title}{Density {Functional} {Theory} {Half}-{Electron} {Self}-{Energy} {Correction} for {Fast} and {Accurate} {Nonadiabatic} {Molecular} {Dynamics}}.
\newblock \emph{\bibinfo{journal}{The Journal of Physical Chemistry Letters}} \textbf{\bibinfo{volume}{12}}, \bibinfo{pages}{10886--10892} (\bibinfo{year}{2021}).

\bibitem{Anisimov_band_1991}
\bibinfo{author}{Anisimov, V.~I.}, \bibinfo{author}{Zaanen, J.} \& \bibinfo{author}{Andersen, O.~K.}
\newblock \bibinfo{title}{Band theory and {Mott} insulators: {Hubbard} {U} instead of {Stoner} {I}}.
\newblock \emph{\bibinfo{journal}{Physical Review B}} \textbf{\bibinfo{volume}{44}}, \bibinfo{pages}{943--954} (\bibinfo{year}{1991}).

\bibitem{hu_choice_2011}
\bibinfo{author}{Hu, Z.} \& \bibinfo{author}{Metiu, H.}
\newblock \bibinfo{title}{Choice of {U} for {DFT}+{U} {Calculations} for {Titanium} {Oxides}}.
\newblock \emph{\bibinfo{journal}{The Journal of Physical Chemistry C}} \textbf{\bibinfo{volume}{115}}, \bibinfo{pages}{5841--5845} (\bibinfo{year}{2011}).

\bibitem{loschen_First-principles_2007}
\bibinfo{author}{Loschen, C.}, \bibinfo{author}{Carrasco, J.}, \bibinfo{author}{Neyman, K.~M.} \& \bibinfo{author}{Illas, F.}
\newblock \bibinfo{title}{First-principles \${\textbackslash}mathrm\{{LDA}\}+{\textbackslash}mathrm\{{U}\}\$ and \${\textbackslash}mathrm\{{GGA}\}+{\textbackslash}mathrm\{{U}\}\$ study of cerium oxides: {Dependence} on the effective {U} parameter}.
\newblock \emph{\bibinfo{journal}{Physical Review B}} \textbf{\bibinfo{volume}{75}}, \bibinfo{pages}{035115} (\bibinfo{year}{2007}).

\bibitem{wang_real_2019}
\bibinfo{author}{Wang, C.-Y.}, \bibinfo{author}{Elliott, P.}, \bibinfo{author}{Sharma, S.} \& \bibinfo{author}{Dewhurst, J.~K.}
\newblock \bibinfo{title}{Real time scissor correction in {TD}-{DFT}}.
\newblock \emph{\bibinfo{journal}{Journal of Physics: Condensed Matter}} \textbf{\bibinfo{volume}{31}}, \bibinfo{pages}{214002} (\bibinfo{year}{2019}).

\bibitem{Cignoni_machine_2023}
\bibinfo{author}{Cignoni, E.}, \bibinfo{author}{Cupellini, L.} \& \bibinfo{author}{Mennucci, B.}
\newblock \bibinfo{title}{Machine {Learning} {Exciton} {Hamiltonians} in {Light}-{Harvesting} {Complexes}}.
\newblock \emph{\bibinfo{journal}{Journal of Chemical Theory and Computation}} \textbf{\bibinfo{volume}{19}}, \bibinfo{pages}{965--977} (\bibinfo{year}{2023}).

\bibitem{wang_all-atom_2022}
\bibinfo{author}{Wang, Z.}, \bibinfo{author}{Dong, J.}, \bibinfo{author}{Qiu, J.} \& \bibinfo{author}{Wang, L.}
\newblock \bibinfo{title}{All-{Atom} {Nonadiabatic} {Dynamics} {Simulation} of {Hybrid} {Graphene} {Nanoribbons} {Based} on {Wannier} {Analysis} and {Machine} {Learning}}.
\newblock \emph{\bibinfo{journal}{ACS Applied Materials \& Interfaces}} \textbf{\bibinfo{volume}{14}}, \bibinfo{pages}{22929--22940} (\bibinfo{year}{2022}).

\bibitem{dral_Molecular_2021}
\bibinfo{author}{Dral, P.~O.} \& \bibinfo{author}{Barbatti, M.}
\newblock \bibinfo{title}{Molecular excited states through a machine learning lens}.
\newblock \emph{\bibinfo{journal}{Nature Reviews Chemistry}} \textbf{\bibinfo{volume}{5}}, \bibinfo{pages}{388--405} (\bibinfo{year}{2021}).

\bibitem{zhang_mlatom_2024}
\bibinfo{author}{Zhang, L.} \emph{et~al.}
\newblock \bibinfo{title}{{MLatom} {Software} {Ecosystem} for {Surface} {Hopping} {Dynamics} in {Python} with {Quantum} {Mechanical} and {Machine} {Learning} {Methods}}.
\newblock \emph{\bibinfo{journal}{Journal of Chemical Theory and Computation}} \textbf{\bibinfo{volume}{20}}, \bibinfo{pages}{5043--5057} (\bibinfo{year}{2024}).

\bibitem{li_machine_2024}
\bibinfo{author}{Li, X.}, \bibinfo{author}{Lubbers, N.}, \bibinfo{author}{Tretiak, S.}, \bibinfo{author}{Barros, K.} \& \bibinfo{author}{Zhang, Y.}
\newblock \bibinfo{title}{Machine {Learning} {Framework} for {Modeling} {Exciton} {Polaritons} in {Molecular} {Materials}}.
\newblock \emph{\bibinfo{journal}{Journal of Chemical Theory and Computation}} \textbf{\bibinfo{volume}{20}}, \bibinfo{pages}{891--901} (\bibinfo{year}{2024}).

\bibitem{habib_machine_2023}
\bibinfo{author}{Habib, A.}, \bibinfo{author}{Lubbers, N.}, \bibinfo{author}{Tretiak, S.} \& \bibinfo{author}{Nebgen, B.}
\newblock \bibinfo{title}{Machine {Learning} {Models} {Capture} {Plasmon} {Dynamics} in {Ag} {Nanoparticles}}.
\newblock \emph{\bibinfo{journal}{The Journal of Physical Chemistry A}} \textbf{\bibinfo{volume}{127}}, \bibinfo{pages}{3768--3778} (\bibinfo{year}{2023}).

\bibitem{hu_inclusion_2018}
\bibinfo{author}{Hu, D.}, \bibinfo{author}{Xie, Y.}, \bibinfo{author}{Li, X.}, \bibinfo{author}{Li, L.} \& \bibinfo{author}{Lan, Z.}
\newblock \bibinfo{title}{Inclusion of {Machine} {Learning} {Kernel} {Ridge} {Regression} {Potential} {Energy} {Surfaces} in {On}-the-{Fly} {Nonadiabatic} {Molecular} {Dynamics} {Simulation}}.
\newblock \emph{\bibinfo{journal}{The Journal of Physical Chemistry Letters}} \textbf{\bibinfo{volume}{9}}, \bibinfo{pages}{2725--2732} (\bibinfo{year}{2018}).

\bibitem{shu_Nonadiabatic_2022}
\bibinfo{author}{Shu, Y.} \emph{et~al.}
\newblock \bibinfo{title}{Nonadiabatic {Dynamics} {Algorithms} with {Only} {Potential} {Energies} and {Gradients}: {Curvature}-{Driven} {Coherent} {Switching} with {Decay} of {Mixing} and {Curvature}-{Driven} {Trajectory} {Surface} {Hopping}}.
\newblock \emph{\bibinfo{journal}{Journal of Chemical Theory and Computation}} \textbf{\bibinfo{volume}{18}}, \bibinfo{pages}{1320--1328} (\bibinfo{year}{2022}).

\bibitem{T_do_Casal_fewest_2021}
\bibinfo{author}{T~do Casal, M.}, \bibinfo{author}{Toldo, J.~M.}, \bibinfo{author}{Pinheiro, M.} \& \bibinfo{author}{Barbatti, M.}
\newblock \bibinfo{title}{Fewest switches surface hopping with {Baeck}-{An} couplings}.
\newblock \emph{\bibinfo{journal}{Open Research Europe}} \textbf{\bibinfo{volume}{1}}, \bibinfo{pages}{49} (\bibinfo{year}{2021}).

\bibitem{Westermayr_combining_2020}
\bibinfo{author}{Westermayr, J.}, \bibinfo{author}{Gastegger, M.} \& \bibinfo{author}{Marquetand, P.}
\newblock \bibinfo{title}{Combining {SchNet} and {SHARC}: {The} {SchNarc} {Machine} {Learning} {Approach} for {Excited}-{State} {Dynamics}}.
\newblock \emph{\bibinfo{journal}{The Journal of Physical Chemistry Letters}}  (\bibinfo{year}{2020}).

\bibitem{li_Automatic_2021}
\bibinfo{author}{Li, J.} \emph{et~al.}
\newblock \bibinfo{title}{Automatic discovery of photoisomerization mechanisms with nanosecond machine learning photodynamics simulations}.
\newblock \emph{\bibinfo{journal}{Chemical Science}} \textbf{\bibinfo{volume}{12}}, \bibinfo{pages}{5302--5314} (\bibinfo{year}{2021}).

\bibitem{Shakiba_Machine-Learned_2024}
\bibinfo{author}{Shakiba, M.} \& \bibinfo{author}{Akimov, A.~V.}
\newblock \bibinfo{title}{Machine-{Learned} {Kohn}–{Sham} {Hamiltonian} {Mapping} for {Nonadiabatic} {Molecular} {Dynamics}}.
\newblock \emph{\bibinfo{journal}{Journal of Chemical Theory and Computation}} \textbf{\bibinfo{volume}{20}}, \bibinfo{pages}{2992--3007} (\bibinfo{year}{2024}).

\bibitem{zhang_Doping-Induced_2021}
\bibinfo{author}{Zhang, Z.}, \bibinfo{author}{Zhang, Y.}, \bibinfo{author}{Wang, J.}, \bibinfo{author}{Xu, J.} \& \bibinfo{author}{Long, R.}
\newblock \bibinfo{title}{Doping-{Induced} {Charge} {Localization} {Suppresses} {Electron}–{Hole} {Recombination} in {Copper} {Zinc} {Tin} {Sulfide}: {Quantum} {Dynamics} {Combined} with {Deep} {Neural} {Networks} {Analysis}}.
\newblock \emph{\bibinfo{journal}{The Journal of Physical Chemistry Letters}} \textbf{\bibinfo{volume}{12}}, \bibinfo{pages}{835--842} (\bibinfo{year}{2021}).

\bibitem{wang_interpolating_2023}
\bibinfo{author}{Wang, B.} \emph{et~al.}
\newblock \bibinfo{title}{Interpolating {Nonadiabatic} {Molecular} {Dynamics} {Hamiltonian} with {Bidirectional} {Long} {Short}-{Term} {Memory} {Networks}}.
\newblock \emph{\bibinfo{journal}{The Journal of Physical Chemistry Letters}} \textbf{\bibinfo{volume}{14}}, \bibinfo{pages}{7092--7099} (\bibinfo{year}{2023}).

\bibitem{gong_general_2023}
\bibinfo{author}{Gong, X.} \emph{et~al.}
\newblock \bibinfo{title}{General framework for {E}(3)-equivariant neural network representation of density functional theory {Hamiltonian}}.
\newblock \emph{\bibinfo{journal}{Nature Communications}} \textbf{\bibinfo{volume}{14}}, \bibinfo{pages}{2848} (\bibinfo{year}{2023}).

\bibitem{Musaelian_learning_2023}
\bibinfo{author}{Musaelian, A.} \emph{et~al.}
\newblock \bibinfo{title}{Learning local equivariant representations for large-scale atomistic dynamics}.
\newblock \emph{\bibinfo{journal}{Nature Communications}} \textbf{\bibinfo{volume}{14}}, \bibinfo{pages}{579} (\bibinfo{year}{2023}).

\bibitem{zhong_transferable_2023}
\bibinfo{author}{Zhong, Y.}, \bibinfo{author}{Yu, H.}, \bibinfo{author}{Su, M.}, \bibinfo{author}{Gong, X.} \& \bibinfo{author}{Xiang, H.}
\newblock \bibinfo{title}{Transferable equivariant graph neural networks for the {Hamiltonians} of molecules and solids}.
\newblock \emph{\bibinfo{journal}{npj Computational Materials}} \textbf{\bibinfo{volume}{9}}, \bibinfo{pages}{1--13} (\bibinfo{year}{2023}).

\bibitem{zhong_Universal_2024}
\bibinfo{author}{Zhong, Y.} \emph{et~al.}
\newblock \bibinfo{title}{Universal {Machine} {Learning} {Kohn}–{Sham} {Hamiltonian} for {Materials}}.
\newblock \emph{\bibinfo{journal}{Chinese Physics Letters}} \textbf{\bibinfo{volume}{41}}, \bibinfo{pages}{077103} (\bibinfo{year}{2024}).

\bibitem{tang_efficient_2023}
\bibinfo{author}{Tang, Z.} \emph{et~al.}
\newblock \bibinfo{title}{Efficient hybrid density functional calculation by deep learning} (\bibinfo{year}{2023}).

\bibitem{perdew_generalized_1996}
\bibinfo{author}{Perdew, J.~P.}, \bibinfo{author}{Burke, K.} \& \bibinfo{author}{Ernzerhof, M.}
\newblock \bibinfo{title}{Generalized {Gradient} {Approximation} {Made} {Simple}}.
\newblock \emph{\bibinfo{journal}{Physical Review Letters}} \textbf{\bibinfo{volume}{77}}, \bibinfo{pages}{3865--3868} (\bibinfo{year}{1996}).

\bibitem{hammes-schiffer_proton_1994}
\bibinfo{author}{Hammes-Schiffer, S.} \& \bibinfo{author}{Tully, J.~C.}
\newblock \bibinfo{title}{Proton transfer in solution: {Molecular} dynamics with quantum transitions}.
\newblock \emph{\bibinfo{journal}{The Journal of Chemical Physics}} \textbf{\bibinfo{volume}{101}}, \bibinfo{pages}{4657--4667} (\bibinfo{year}{1994}).

\bibitem{tully_molecular_1990}
\bibinfo{author}{Tully, J.~C.}
\newblock \bibinfo{title}{Molecular dynamics with electronic transitions}.
\newblock \emph{\bibinfo{journal}{The Journal of Chemical Physics}} \textbf{\bibinfo{volume}{93}}, \bibinfo{pages}{1061--1071} (\bibinfo{year}{1990}).

\bibitem{jaeger_Decoherence-induced_2012}
\bibinfo{author}{Jaeger, H.~M.}, \bibinfo{author}{Fischer, S.} \& \bibinfo{author}{Prezhdo, O.~V.}
\newblock \bibinfo{title}{Decoherence-induced surface hopping}.
\newblock \emph{\bibinfo{journal}{The Journal of Chemical Physics}} \textbf{\bibinfo{volume}{137}}, \bibinfo{pages}{22A545} (\bibinfo{year}{2012}).

\bibitem{akimov_pyxaid_2013}
\bibinfo{author}{Akimov, A.~V.} \& \bibinfo{author}{Prezhdo, O.~V.}
\newblock \bibinfo{title}{The {PYXAID} {Program} for {Non}-{Adiabatic} {Molecular} {Dynamics} in {Condensed} {Matter} {Systems}}.
\newblock \emph{\bibinfo{journal}{Journal of Chemical Theory and Computation}} \textbf{\bibinfo{volume}{9}}, \bibinfo{pages}{4959--4972} (\bibinfo{year}{2013}).

\bibitem{wang_mixed_2011}
\bibinfo{author}{Wang, L.}, \bibinfo{author}{Beljonne, D.}, \bibinfo{author}{Chen, L.} \& \bibinfo{author}{Shi, Q.}
\newblock \bibinfo{title}{Mixed quantum-classical simulations of charge transport in organic materials: {Numerical} benchmark of the {Su}-{Schrieffer}-{Heeger} model}.
\newblock \emph{\bibinfo{journal}{The Journal of Chemical Physics}} \textbf{\bibinfo{volume}{134}}, \bibinfo{pages}{244116} (\bibinfo{year}{2011}).

\bibitem{ORegan_low-cost_1991}
\bibinfo{author}{O'Regan, B.} \& \bibinfo{author}{Grätzel, M.}
\newblock \bibinfo{title}{A low-cost, high-efficiency solar cell based on dye-sensitized colloidal {TiO2} films}.
\newblock \emph{\bibinfo{journal}{Nature}} \textbf{\bibinfo{volume}{353}}, \bibinfo{pages}{737--740} (\bibinfo{year}{1991}).

\bibitem{Linsebigler_Photocatalysis_1995}
\bibinfo{author}{Linsebigler, A.~L.}, \bibinfo{author}{Lu, G.} \& \bibinfo{author}{Yates, J. T.~J.}
\newblock \bibinfo{title}{Photocatalysis on {TiO2} {Surfaces}: {Principles}, {Mechanisms}, and {Selected} {Results}}.
\newblock \emph{\bibinfo{journal}{Chemical Reviews}} \textbf{\bibinfo{volume}{95}}, \bibinfo{pages}{735--758} (\bibinfo{year}{1995}).

\bibitem{Nakata_tio2_2012}
\bibinfo{author}{Nakata, K.} \& \bibinfo{author}{Fujishima, A.}
\newblock \bibinfo{title}{{TiO2} photocatalysis: {Design} and applications}.
\newblock \emph{\bibinfo{journal}{Journal of Photochemistry and Photobiology C: Photochemistry Reviews}} \textbf{\bibinfo{volume}{13}}, \bibinfo{pages}{169--189} (\bibinfo{year}{2012}).

\bibitem{akimov_theoretical_2013}
\bibinfo{author}{Akimov, A.~V.}, \bibinfo{author}{Neukirch, A.~J.} \& \bibinfo{author}{Prezhdo, O.~V.}
\newblock \bibinfo{title}{Theoretical {Insights} into {Photoinduced} {Charge} {Transfer} and {Catalysis} at {Oxide} {Interfaces}}.
\newblock \emph{\bibinfo{journal}{Chemical Reviews}} \textbf{\bibinfo{volume}{113}}, \bibinfo{pages}{4496--4565} (\bibinfo{year}{2013}).

\bibitem{chu_ultrafast_2016}
\bibinfo{author}{Chu, W.} \emph{et~al.}
\newblock \bibinfo{title}{Ultrafast {Dynamics} of {Photongenerated} {Holes} at a {CH3OH}/{TiO2} {Rutile} {Interface}}.
\newblock \emph{\bibinfo{journal}{Journal of the American Chemical Society}} \textbf{\bibinfo{volume}{138}}, \bibinfo{pages}{13740--13749} (\bibinfo{year}{2016}).

\bibitem{you_Correlated_2024}
\bibinfo{author}{You, P.} \emph{et~al.}
\newblock \bibinfo{title}{Correlated electron–nuclear dynamics of photoinduced water dissociation on rutile {TiO2}}.
\newblock \emph{\bibinfo{journal}{Nature Materials}} \bibinfo{pages}{1--7} (\bibinfo{year}{2024}).

\bibitem{Amtout_optical_1995}
\bibinfo{author}{Amtout, A.} \& \bibinfo{author}{Leonelli, R.}
\newblock \bibinfo{title}{Optical properties of rutile near its fundamental band gap}.
\newblock \emph{\bibinfo{journal}{Physical Review B}} \textbf{\bibinfo{volume}{51}}, \bibinfo{pages}{6842--6851} (\bibinfo{year}{1995}).

\bibitem{Blakemore_Semiconducting_1982}
\bibinfo{author}{Blakemore, J.~S.}
\newblock \bibinfo{title}{Semiconducting and other major properties of gallium arsenide}.
\newblock \emph{\bibinfo{journal}{Journal of Applied Physics}} \textbf{\bibinfo{volume}{53}}, \bibinfo{pages}{R123--R181} (\bibinfo{year}{1982}).

\bibitem{landmann_electronic_2012}
\bibinfo{author}{Landmann, M.}, \bibinfo{author}{Rauls, E.} \& \bibinfo{author}{Schmidt, W.~G.}
\newblock \bibinfo{title}{The electronic structure and optical response of rutile, anatase and brookite {TiO2}}.
\newblock \emph{\bibinfo{journal}{Journal of Physics: Condensed Matter}} \textbf{\bibinfo{volume}{24}}, \bibinfo{pages}{195503} (\bibinfo{year}{2012}).

\bibitem{si}
 \bibinfo{note}{See Supplemental Information for more information.}

\bibitem{qin_honpas_2015}
\bibinfo{author}{Qin, X.}, \bibinfo{author}{Shang, H.}, \bibinfo{author}{Xiang, H.}, \bibinfo{author}{Li, Z.} \& \bibinfo{author}{Yang, J.}
\newblock \bibinfo{title}{{HONPAS}: {A} linear scaling open-source solution for large system simulations}.
\newblock \emph{\bibinfo{journal}{International Journal of Quantum Chemistry}} \textbf{\bibinfo{volume}{115}}, \bibinfo{pages}{647--655} (\bibinfo{year}{2015}).

\bibitem{heyd_hybrid_2003}
\bibinfo{author}{Heyd, J.}, \bibinfo{author}{Scuseria, G.~E.} \& \bibinfo{author}{Ernzerhof, M.}
\newblock \bibinfo{title}{Hybrid functionals based on a screened {Coulomb} potential}.
\newblock \emph{\bibinfo{journal}{The Journal of Chemical Physics}} \textbf{\bibinfo{volume}{118}}, \bibinfo{pages}{8207--8215} (\bibinfo{year}{2003}).

\bibitem{chu_low-frequency_2020}
\bibinfo{author}{Chu, W.}, \bibinfo{author}{Zheng, Q.}, \bibinfo{author}{Prezhdo, O.~V.}, \bibinfo{author}{Zhao, J.} \& \bibinfo{author}{Saidi, W.~A.}
\newblock \bibinfo{title}{Low-frequency lattice phonons in halide perovskites explain high defect tolerance toward electron-hole recombination}.
\newblock \emph{\bibinfo{journal}{Science Advances}} \textbf{\bibinfo{volume}{6}}, \bibinfo{pages}{eaaw7453} (\bibinfo{year}{2020}).

\bibitem{wang_Effective_2022}
\bibinfo{author}{Wang, S.} \emph{et~al.}
\newblock \bibinfo{title}{Effective lifetime of non-equilibrium carriers in semiconductors from non-adiabatic molecular dynamics simulations}.
\newblock \emph{\bibinfo{journal}{Nature Computational Science}} \textbf{\bibinfo{volume}{2}}, \bibinfo{pages}{486--493} (\bibinfo{year}{2022}).

\bibitem{Shakiba_Dependence_2023}
\bibinfo{author}{Shakiba, M.} \& \bibinfo{author}{Akimov, A.~V.}
\newblock \bibinfo{title}{Dependence of {Electron}–{Hole} {Recombination} {Rates} on {Charge} {Carrier} {Concentration}: {A} {Case} {Study} of {Nonadiabatic} {Molecular} {Dynamics} in {Graphitic} {Carbon} {Nitride} {Monolayers}}.
\newblock \emph{\bibinfo{journal}{The Journal of Physical Chemistry C}} \textbf{\bibinfo{volume}{127}}, \bibinfo{pages}{9083--9096} (\bibinfo{year}{2023}).

\bibitem{janotti_hybrid_2010}
\bibinfo{author}{Janotti, A.} \emph{et~al.}
\newblock \bibinfo{title}{Hybrid functional studies of the oxygen vacancy in {TiO} 2}.
\newblock \emph{\bibinfo{journal}{Physical Review B}} \textbf{\bibinfo{volume}{81}}, \bibinfo{pages}{085212} (\bibinfo{year}{2010}).

\bibitem{zhang_Origin_2022}
\bibinfo{author}{Zhang, X.} \& \bibinfo{author}{Wei, S.-H.}
\newblock \bibinfo{title}{Origin of {Efficiency} {Enhancement} by {Lattice} {Expansion} in {Hybrid}-{Perovskite} {Solar} {Cells}}.
\newblock \emph{\bibinfo{journal}{Physical Review Letters}} \textbf{\bibinfo{volume}{128}}, \bibinfo{pages}{136401} (\bibinfo{year}{2022}).

\bibitem{maity_study_2018}
\bibinfo{author}{Maity, P.}, \bibinfo{author}{Mohammed, O.~F.}, \bibinfo{author}{Katsiev, K.} \& \bibinfo{author}{Idriss, H.}
\newblock \bibinfo{title}{Study of the {Bulk} {Charge} {Carrier} {Dynamics} in {Anatase} and {Rutile} {TiO2} {Single} {Crystals} by {Femtosecond} {Time}-{Resolved} {Spectroscopy}}.
\newblock \emph{\bibinfo{journal}{The Journal of Physical Chemistry C}} \textbf{\bibinfo{volume}{122}}, \bibinfo{pages}{8925--8932} (\bibinfo{year}{2018}).

\bibitem{Furube_charge_1999}
\bibinfo{author}{Furube, A.}, \bibinfo{author}{Asahi, T.}, \bibinfo{author}{Masuhara, H.}, \bibinfo{author}{Yamashita, H.} \& \bibinfo{author}{Anpo, M.}
\newblock \bibinfo{title}{Charge {Carrier} {Dynamics} of {Standard} {TiO2} {Catalysts} {Revealed} by {Femtosecond} {Diffuse} {Reflectance} {Spectroscopy}}.
\newblock \emph{\bibinfo{journal}{The Journal of Physical Chemistry B}} \textbf{\bibinfo{volume}{103}}, \bibinfo{pages}{3120--3127} (\bibinfo{year}{1999}).

\bibitem{yamada_determination_2012}
\bibinfo{author}{Yamada, Y.} \& \bibinfo{author}{Kanemitsu, Y.}
\newblock \bibinfo{title}{Determination of electron and hole lifetimes of rutile and anatase {TiO2} single crystals}.
\newblock \emph{\bibinfo{journal}{Applied Physics Letters}} \textbf{\bibinfo{volume}{101}}, \bibinfo{pages}{133907} (\bibinfo{year}{2012}).

\bibitem{zhong_Accelerating_2023}
\bibinfo{author}{Zhong, Y.}, \bibinfo{author}{Tao, Z.}, \bibinfo{author}{Chu, W.}, \bibinfo{author}{Gong, X.} \& \bibinfo{author}{Xiang, H.}
\newblock \bibinfo{title}{Accelerating the calculation of electron-phonon coupling by machine learning methods} (\bibinfo{year}{2023}).
\newblock \bibinfo{note}{Preprint at http://arxiv.org/abs/2302.00439}.

\bibitem{zheng_Spin-orbit_2022}
\bibinfo{author}{Zheng, Z.}, \bibinfo{author}{Zheng, Q.} \& \bibinfo{author}{Zhao, J.}
\newblock \bibinfo{title}{Spin-orbit coupling induced demagnetization in {Ni}: \textit{{Ab} initio} nonadiabatic molecular dynamics perspective}.
\newblock \emph{\bibinfo{journal}{Physical Review B}} \textbf{\bibinfo{volume}{105}}, \bibinfo{pages}{085142} (\bibinfo{year}{2022}).

\bibitem{zhong_Accelerating_2023-1}
\bibinfo{author}{Zhong, Y.}, \bibinfo{author}{Zhang, B.}, \bibinfo{author}{Yu, H.}, \bibinfo{author}{Gong, X.} \& \bibinfo{author}{Xiang, H.}
\newblock \bibinfo{title}{Accelerating the electronic-structure calculation of magnetic systems by equivariant neural networks} (\bibinfo{year}{2023}).
\newblock \bibinfo{note}{Preprint at http://arxiv.org/abs/2306.01558}.

\bibitem{ozaki_Variationally_2003}
\bibinfo{author}{Ozaki, T.}
\newblock \bibinfo{title}{Variationally optimized atomic orbitals for large-scale electronic structures}.
\newblock \emph{\bibinfo{journal}{Physical Review B}} \textbf{\bibinfo{volume}{67}}, \bibinfo{pages}{155108} (\bibinfo{year}{2003}).

\bibitem{GRANT_properties_1959}
\bibinfo{author}{GRANT, F.~A.}
\newblock \bibinfo{title}{Properties of {Rutile} ({Titanium} {Dioxide})}.
\newblock \emph{\bibinfo{journal}{Reviews of Modern Physics}} \textbf{\bibinfo{volume}{31}}, \bibinfo{pages}{646--674} (\bibinfo{year}{1959}).

\bibitem{akimov_simple_2018}
\bibinfo{author}{Akimov, A.~V.}
\newblock \bibinfo{title}{A {Simple} {Phase} {Correction} {Makes} a {Big} {Difference} in {Nonadiabatic} {Molecular} {Dynamics}}.
\newblock \emph{\bibinfo{journal}{The Journal of Physical Chemistry Letters}} \textbf{\bibinfo{volume}{9}}, \bibinfo{pages}{6096--6102} (\bibinfo{year}{2018}).

\end{thebibliography}

\section*{Author contributions}
H.J.X. and W.B.C. proposed the research and the methodology in this work. C.W.Z. and Y.Z. wrote the codes, performed the NAMD calculations, and wrote the manuscript. H.J.X., W.B.C., X.G.G, Z.G.L., and O.V.P. revised the manuscript. All authors discussed the results.

\section*{Competing interests}
The authors declare no competing interests.

\end{document}